\documentclass[amsmath,amssymb,twocolumn]{revtex4}
\usepackage{subfig}
\usepackage{dcolumn}
\usepackage{bm}
\usepackage{amsmath,mathrsfs}
\usepackage{graphicx,epsfig}
\usepackage{hyperref}
\usepackage{epsf}
\usepackage{color}

\begin{document}


\title{{\bf Wormholes in $f(Q,T)$ gravity with different matter Lagrangian density}}

\author{S. Nasirimoghadam}\email{soudabe.nasirimoghadam@gmail.com}
\author{ F. Parsaei }\email{fparsaei@gmail.com}
\author{S. Rastgoo}\email{rastgoo@sirjantech.ac.ir}

\affiliation{ Physics Department , Sirjan University of Technology, Sirjan 78137, Iran.}



\begin{abstract}
\par    This study explores asymptotically flat wormhole solutions in $f(Q,T)=\alpha Q+ \beta T$ gravity, expanding upon our prior work (arXiv:2602.00527v1) with matter Lagrangian density, $L_m=-P$ . Here, we examine the implications of employing $L_m=-T$ and $L_m=\rho$. The field equations, derived via action variation, share a common general structure but are fundamentally dictated by the parameters $\alpha$ and $\beta$ through the coefficients  $A_i$. Solutions with linear and asymptotically linear equation of state are explored. We conclude that non-exotic asymptotically flat wormhole solutions exist for all considered matter Lagrangian densities. A key outcome is the demonstration that different $L_m$ choices enable the same shape function to be supported by varied fluid configurations, or vice versa, identical fluids to yield different geometries. The energy conditions and physical characteristics of these solutions are shown to be distinct and critically dependent on the selected $L_m$. \\
\end{abstract}

\maketitle
\section{Introduction}

A primary challenge in the field of wormhole physics is the necessity for the throat to be upheld by matter that contravenes classical energy conditions (ECs), especially the null energy condition (NEC) \cite{Visser}. This type of matter, often termed exotic matter, exhibits characteristics that are not found in typical classical fluids, leading to questions regarding its physical feasibility and stability. As a result, a significant area of investigation within wormhole theory has concentrated on reducing the quantity of
exotic matter needed or substituting it with sources that are physically justifiable. One important strategy is to confine the violation of the ECs to an arbitrarily small region around the throat \cite{cut, cut2, cut3, variable, foad}. An alternative approach investigates wormholes that are sustained by feasible exotic sources which emerge naturally within the realms of quantum field theory or cosmology. A notable instance includes phantom scalar fields characterized by negative kinetic energy that can effectively contravene the NEC and maintain what are referred to as phantom wormholes \cite{phantom, phantom2, phantom1}. Similarly, wormhole solutions that are backed by Casimir-type energy configurations \cite{casm} indicate that quantum vacuum fluctuations could provide a more plausible source of exoticity, particularly at microscopic scales.

In recent times, revised theories of gravity have offered a different perspective: rather than manually introducing exotic matter, one can ascribe the effective violation of the NEC to higher order curvature terms or non-minimal couplings within the gravitational sector. Within this framework, conventional matter can comply with the standard ECs, while the effective stress-energy-tensor resulting from geometric corrections serves as the exotic source. Braneworld \cite{b1, b2, b3}, Born-Infeld theory \cite{Bo, Bo1}, quadratic gravity \cite{quad, quad1}, Einstein-Cartan gravity \cite{Cartan, Cartan1, Cartan2, Cartan3}, Rastall gravity \cite{Rast, Rast1, Sadegh}, $f(Q)$ gravity \cite{fq, fq1, fq2, fq4, fq44, fq5}, $f(R)$ gravity \cite{Nojiri, fR11, fR22, fR55}, $f(T,\mathcal{T})$ gravity \cite{Must1, Err, Riz, Must2, foad3}, $f(R, l_m)$ gravity \cite{L1, L2, L7, L10, L12, SR2}, $f(R,T)$ gravity \cite{Azizi, Moa, Zub, Zub2, fr2, Sha, Ban, Sarkar, SR1, foad4} and $f(Q,T)$ gravity \cite{1, 11, 12a,12, 2, 3, 4, 5, 6, 7, 8, 9, 10, 13, 14, fq-sara}  are some
example. Generally, Lagrangian-based theories of gravity are founded on a variational principle:
an action functional ($S[g_{\mu \nu}, \psi]$) is formulated, which is contingent upon the metric (and potentially other fields
($\psi$)), and it is required that the action remains stationary when subjected to arbitrary variations of the dynamical variables. In this realm, the matter Lagrangian
density ($L_m$) is fundamental in the development of gravitational theories, as it encapsulates the physical characteristics of matter fields and dictates their interaction
with the geometry of spacetime. In the context of general relativity (GR), various forms of $L_m$ that represent the same physical scenario typically yield equivalent field
equations, making the specific selection of the matter Lagrangian largely irrelevant at the classical level. Nevertheless, this equivalence is not maintained in modified
gravity theories that incorporate non-minimal couplings between matter and geometric entities, such as $f(R, Lm)$ , $f(R, T )$, $f(T,\mathcal{T})$, and $f(Q,T)$ gravity. Within these theoretical frameworks, the gravitational action is explicitly dependent on $L_m$ or on quantities derived from it, thus rendering the choice of the matter Lagrangian density a significant physical assumption rather than merely a technical formality.

Recently, we have conducted an investigation into wormholes within the framework of $f(Q, T)$ gravity \cite{fq-sara}. The
$f(Q, T)$ gravity represents a modified gravitational theory that broadens the symmetric teleparallel formulation
of GR \cite{fqt}. In this context, gravity is characterized not by curvature (as seen in GR) or torsion (as described in Einstein-Cartan theory), but rather by non-metricity the
quantification of how the gravitational field fails to maintain lengths during parallel transport. The primary geometric quantity is the non-metricity scalar $Q$, which is
analogous to the Ricci scalar $R$ found in GR. The gravitational Lagrangian is an arbitrary function of both $Q$ and the trace $T$ of the matter energy-momentum tensor (EMT). This relationship between geometry ($Q$) and matter ($T$ ) signifies a non-minimal interaction, implying that the matter content has a direct impact on the gravitational
field equations beyond the conventional source term. As a result, $f(Q,T )$ gravity forecasts deviations from GR in both weak and strong gravitational regimes and is currently being explored as a possible alternative to dark energy for elucidating cosmic acceleration, as well as for tackling issues such as galaxy rotation curves and primordial fluctuations.

In \cite{fq-sara}, we have adopted $L_m =-P$ to find non-exotic asymptotically flat wormhole solutions in $f(Q,T)$ gravity. In the context of fluid descriptions frequently employed in astrophysical and cosmological studies, various inequivalent representations of $L_m$ have been introduced in the literature. These include, along with more comprehensive
formulations that are based on particle number density or thermodynamic potentials. Each selection embodies different foundational assumptions regarding the microscopic composition of matter, resulting in unique EMT when they are non-minimally coupled to geometry. Consequently, observable predictions—such as equilibrium conditions, stability criteria, and particle trajectories may differ based on the chosen representation of $L_m$. Therefore, comprehending and meticulously justifying the selection of the matter Lagrangian density is crucial when examining solutions and their physical implications in modified gravity theories, especially in the investigation of compact objects, wormholes, and cosmological dynamics. The selection of the matter Lagrangian for perfect fluids has been previously addressed in the literature. While $L_m=-p$ and $L_m=\rho$ are the most frequently utilized options, the choice of $L_m=-T$ has also been encountered \cite{20}.
 In reference \cite{Haghani}, a study concerning the issue of the second variation of the perfect fluid matter Lagrangian in relation to the components of the metric tensor, along with an analysis of its effects on modified gravity theories, is provided. Within the context of the $f (R, L_m)$ theory, it was demonstrated in \cite{21} that if one adopts the expression $L_m = -p$ for the Lagrangian density, where $p$ represents the pressure, the additional force disappears in the case of dust. Conversely, when considering the form $L_m = \rho$ for the matter Lagrangian, the extra force does not disappear \cite{22}. In \cite{23}, it has been demonstrated that the matter Lagrangian holds significant importance in the $f(R, T)$ theory of gravity. In \cite{24}, the authors asserted that the matter Lagrangian represents its total energy density rather than its pressure. Thermodynamic invariance of the EMT under matter Lagrangian choices and its astrophysical implications in $f(R,T)$ gravity are studied in  \cite{25}. Bhagat et al. have investigated the effects of matter Lagrangian in $f(Q,T)$ gravity \cite{26}. In \cite{26b}, the authors illustrated that when the fluid comprises localized energy concentrations with a constant rest mass and structure (solitons), the average on-shell Lagrangian of a perfect fluid is expressed as $L_m = -T$.

In this paper, we examine the consequences of various $L_m$ in $f(Q,T)$ gravity concerning the development and evaluation of wormholes. We will investigate how
this altered theory can produce feasible wormhole solutions that comply with ECs and assess the capacity of $f(Q, T)$ gravity to enhance our comprehension of certain
unresolved issues in GR. Through this exploration, we intend to illuminate the significance of $L_m$ in a Lagrangian-based modified gravity theory ($f(Q,T)$) and the fascinating notion of traversable wormholes.
The manuscript is organized as follows. Section \ref{sec2} introduces the wormhole concept. This is followed by a brief overview of the $f(Q,T)$ theory and the associated
classical ECs. In Section \ref{sec3}, we present solutions that satisfy these ECs for three distinct matter Lagrangian densities. Section \ref{sec4a} provides a brief discussion on wormholes within the framework of non-linear $f(Q,T)$ gravity. The resulting solutions are then compared in Section \ref{sec4}. Finally, our main conclusions are summarized in Section \ref{sec5}. Throughout, we work in geometrized
units where $c = 8 \pi G = 1$.

\section{Basic formalism} \label{sec2}
In this section, we provide a concise overview of the fundamental structure of the wormhole and the formulation of $f(Q,T)$ gravity. The approach is akin to that in \cite{fq-sara}. For a spherically symmetric and static wormhole, the metric is expressed as follows
\begin{equation}\label{1}
ds^2=-U(r)dt^2+({1-\frac{b(r)}{r}})^{-1}{dr^2}+r^2(d\theta^2+\sin^2\theta.
d\phi^2)
\end{equation}
In this context, $U(r)=\exp(2\phi(r)$) and $\phi(r)$ denotes the redshift function which is related to gravitational redshift. In addition, $b(r)$ is the shape function, which influences the curvature and geometry of the wormhole, directly. Two different Universes or separate regions within the same Universe is connected by the throat of a wormhole ($r_0$). The necessary condition at throat is
\begin{equation}\label{2}
b(r_0)=r_0.
\end{equation}
 For a traversable wormhole, the shape function must fulfill the following conditions
\begin{equation}\label{3}
b'(r_0)<1,
\end{equation}
and
\begin{equation}\label{4}
b(r)<r,\ \ {\rm for} \ \ r>r_0.
\end{equation}
Furthermore, the condition of being asymptotically flat must be fulfilled on a large cosmic scale,
\begin{equation}\label{5}
\lim_{r\rightarrow \infty}\frac{b(r)}{r}=0,\qquad   \lim_{r\rightarrow \infty}U(r)=1.
\end{equation}

To prevent the formations of horizons in the traversable wormholes, the redshift function $\phi(r)$ must be finite at all points. Consequently, a straightforward expression for $\phi(r)$ is $\phi(r)=constant$. In this paper, we examine a constant redshift function. The use of a constant redshift function guarantees the absence of tidal forces.
It is important to highlight that we have taken into account an anisotropic fluid with the EMT represented as $T_{\nu}^{\mu}=diag[-\rho,p,p_{t},p_{t}]$, where $\rho$, $p$, and $p_{t}$ denote the energy density, radial pressure, and tangential pressure, respectively. It should be noted that $T=-\rho+p+2p_{t}$ is the trace of EMT and $P=\frac{p+2p_{t}}{3}$.

Now, it is useful to provide a concise overview of the $f(Q,T)$ of gravity.
In this theory, the total action is represented by
\begin{equation}\label{6a}
\mathcal{S}=\int\frac{1}{16\pi}\,f(Q,T)\sqrt{-g}\,d^4x+\int \mathcal{L}_m\,\sqrt{-g}\,d^4x\, ,
\end{equation}
where $f(Q,T)$ is a general function of the non-metricity scalar, $Q$, and the trace of EMT, $T$. The $\sqrt{-g}$ and $L_{m}$ denote the determinant of the metric, and the matter Lagrangian density, respectively \cite{6}.
The non-metricity tensor is formulated as\cite{6}
\begin{equation}\label{6ab}
Q_{\lambda\mu\nu}=\bigtriangledown_{\lambda} g_{\mu\nu}.
\end{equation}
Another important component of this theory is the non-metricity conjugate, also known as the superpotential, which takes the following form:
\begin{align}\label{6ab1}
P^\alpha\;_{\mu\nu}=&\frac{1}{4}\left[-Q^\alpha\;_{\mu\nu}+2Q_{(\mu}\;^\alpha\;_{\nu)}+Q^\alpha g_{\mu\nu}-\tilde{Q}^\alpha g_{\mu\nu}\right.\nonumber\\&\left.-\delta^\alpha_{(\mu}Q_{\nu)}\right],
\end{align}
in which
\begin{equation}
\label{a1}
Q_{\alpha}=Q_{\alpha}\;^{\mu}\;_{\mu},\; \tilde{Q}_\alpha=Q^\mu\;_{\alpha\mu}
\end{equation}
denote the traces of the non-metricity tensor. Also, the non-metricity scalar $Q$ and disformation $L^\beta_{\,\,\,\mu\nu}$ can be demonstrated as \cite{6}
\begin{eqnarray}
\label{a2}
Q &=& -Q_{\alpha\mu\nu}\,P^{\alpha\mu\nu}\nonumber\\
&=& -g^{\mu\nu}\left(L^\beta_{\,\,\,\alpha\mu}\,L^\alpha_{\,\,\,\nu\beta}-L^\beta_{\,\,\,\alpha\beta}\,L^\alpha_{\,\,\,\mu\nu}\right),
\end{eqnarray}
and
\begin{equation}\label{a3}
L^\beta_{\,\,\,\mu\nu}=\frac{1}{2}Q^\beta_{\,\,\,\mu\nu}-Q_{(\mu\,\,\,\,\,\,\nu)}^{\,\,\,\,\,\,\beta}.
\end{equation}
The variation of the action concerning the metric tensor leads to the generation of the field equations as follows
\begin{align}\label{6b}
\frac{-2}{\sqrt{-g}}&\bigtriangledown_\alpha\left(\sqrt{-g}\,f_Q\,P^\alpha\;_{\mu\nu}\right)-\frac{1}{2}g_{\mu\nu}f
+f_T \left(T_{\mu\nu} +\Theta_{\mu\nu}\right)\nonumber\\
-&f_Q\left(P_{\mu\alpha\beta}\,Q_\nu\;^{\alpha\beta}-2\,Q^
{\alpha\beta}\,\,_{\mu}\,P_{\alpha\beta\nu}\right)=8\pi T_{\mu\nu},
\end{align}
where, $f_{Q}=\frac{\partial f}{\partial Q}$ and $f_{T}=\frac{\partial f}{\partial T}$.
The EMT  can be represented as
\begin{equation}\label{a4}
T_{\mu\nu}=-\frac{2}{\sqrt{-g}}\frac{\delta\left(\sqrt{-g}\,\mathcal{L}_m\right)}{\delta g^{\mu\nu}},
\end{equation}
and
\begin{equation}\label{a5}
\Theta_{\mu\nu}=g^{\alpha\beta}\frac{\delta T_{\alpha\beta}}{\delta g^{\mu\nu}}.
\end{equation}
The non-metricity scalar can be derived for the metric \eqref{1} as
\begin{equation}\label{a6}
Q=-\frac{b}{r}\left[\frac{rb'-b}{r(r-b)}+2\phi'\right].
\end{equation}
Considering all the aforementioned explanations, the components of the field equations can typically be expressed in relation to $L_m$ as follows:
\begin{align}\label{3e}
\rho=\frac{(r-b)}{2r^3}&\left[f_{\text{Q}}\left(\frac{(2r-b)(rb'-b)}{(r-b)^2}+\frac{2b}{(r-b)}\right)\right.\nonumber\\+\frac{fr^3}{r-b}
&\left.-\frac{2r^3f_T(\rho-L_m)}{(r-b)}+\frac{2brf_{\text{QQ}}Q'}{(r-b)}\right],
\end{align}
\begin{align}\label{3f}
p_r=\frac{(r-b)}{2r^3}&\left[-\frac{b}{r-b}f_{\text{Q}}\left(\frac{rb'-b}{r-b}+2\right)-\frac{fr^3}{r-b}\right.\nonumber\\
&\left.-\frac{2r^3f_T(L_m+p_r)}{r-b}-\frac{2brf_{\text{QQ}}Q'}{r-b}\right],
\end{align}
\begin{align}\label{3g}
p_t=-\frac{(r-b)}{4r^2}&\left[f_{\text{Q}} \left(\frac{2(rb'-b)}{(r-b)^2}\right)+\frac{2fr^2}{r-b}\right.\nonumber\\
&\left.\;\;\;\;\;\;\;\;\;+\frac{4r^2f_T(L_m+p_t)}{(r-b)}\right].
\end{align}
It is clear that the field equations rely on the configuration of the function $f(Q,T)$. By employing these specific field equations, one can investigate various wormhole solutions within the framework of the $f(Q,T)$ gravity model.

The conventional ECs, derived from the Raychaudhuri equations, are employed to examine physically plausible matter configurations. ECs play a crucial role in comprehending gravity and the behavior of matter in the Universe. These conditions establish a set of limitations on the stress-energy-tensor, which represents the distributions of matter and energy across spacetime. The four conditions that constitute this framework are expressed as follows
\begin{eqnarray}\label{21}
\textbf{NEC}&:& \rho+p_r\geq 0,\quad \rho+p_t\geq 0 \\
\label{21a}
\textbf{WEC}&:& \rho\geq 0, \rho+p_r\geq 0,\quad \rho+p_t\geq 0, \\
\textbf{DEC}&:& \rho\geq 0, \rho-|p_r|\geq 0,\quad \rho-|p_t|\geq 0, \\
\textbf{SEC}&:& \rho+p_r\geq 0,\, \rho+p_t\geq 0,\rho+p_r+2p_t \geq 0.
\end{eqnarray}
In the framework of GR, the NEC is viewed as the most important of these conditions, due to its connection with the energy density. A violation of the NEC in the region close to the throat of a wormhole would indicate the presence of exotic matter with negative energy density, which is not available in standard sources of matter.
Based on reference \cite{fq}, we express the ECs by defining the following functions:
\begin{eqnarray}\label{22}
 H(r)&=& \rho+p_r,\, H_1(r)= \rho+p_t,\, H_2(r)= \rho-|p_r|, \nonumber \\
 H_3(r)&=&\rho-|p_t|,\, H_4(r)= \rho+p_r+2p_t .
\end{eqnarray}
There are various methods for analyzing the asymptotically flat wormhole solutions in the GR framework. In the following Section, we will implement these techniques to locate asymptotically flat wormhole solutions. It is noteworthy that we have taken $r_{0}=1$ as a constant assumption throughout this investigation.

\section{Non-exotic wormhole solutions for different $L_m$}\label{sec3}
In this research, we employ the linear representation of the function $f(Q,T)$ as detailed below:
\begin{equation}\label{99a}
f(Q, T)=\alpha  Q+\beta T,
\end{equation}
where $\alpha$ and $\beta$ are free parameters. By employing the function (\ref{99a}) in Eqs.(\ref{3e})-(\ref{3g}), the  following field equations can be derived
\begin{equation}\label{19aa}
\rho=\frac{\alpha rb'}{(1-\beta)(1+\beta)r^3}+\frac{\beta}{1-\beta}L_m,
\end{equation}
\begin{equation}\label{20aa}
p_r=-\frac{\alpha\left(\beta(rb'-b)+b\right)}{(1-\beta)(1+\beta)r^3}-\frac{\beta}{(1-\beta)}L_m,
\end{equation}
\begin{equation}
p_t=-\frac{\alpha\left[(rb'-b)+\beta(rb'+b)\right]}{2(1-\beta)(1+\beta)r^3}-\frac{\beta}{(1-\beta)}L_m.
\end{equation}
We can easily show that
\begin{eqnarray}\label{11cc}
H(r)&=&\rho(r)+p_r(r)\nonumber \\
&=&\gamma(\alpha,\beta) \frac{(rb'-b)}{r^3},
\end{eqnarray}
and
\begin{eqnarray}\label{11d}
H_1(r)&=&\rho(r)+p_t(r)\nonumber \\
&=&\gamma(\alpha,\beta) \frac{(rb'+b)}{2r^3},
\end{eqnarray}
where
\begin{equation}\label{2a}
\gamma(\alpha,\beta)=\frac{\alpha}{1+\beta}.
\end{equation}
Equations (\ref{11cc}) and (\ref{11d}) imply that
\begin{eqnarray}\label{1cc}
H^f&=&\gamma(\alpha,\beta)H^G,\nonumber \\
H_1^f&=&\gamma(\alpha,\beta)H_1^G,
\end{eqnarray}
where the indexes $f$ and $G$ refer to the $f(Q,T)$ gravity and the GR formalism, respectively. This outcome is consistent with our previous achievements for $L_m=-P$, as referenced in \cite{fq-sara}. It can be inferred that the values of $H$ and $H_1$ presented in (\ref{1cc}) remain unaffected by the choice of $L_m$. As a result, it is clear that solutions which concurrently breach both radial and lateral NEC in the framework of GR can still comply with the NEC in $f(Q,T)$ gravity, provided that the following condition is satisfied:
 \begin{equation}\label{a222}
\gamma(\alpha,\beta)<0.
\end{equation}
In this article, we will employ the following ranges for $\alpha$ and $\beta$, in which $\gamma$ is negative
\begin{equation}\label{cc1}
\alpha>0,\quad\beta<-1
\end{equation}
or
\begin{equation}\label{cc2}
\alpha<0,\quad\beta>-1.
\end{equation}
In earlier research \cite{fq-sara}, the field equations pertaining to the scenario where $L_m=-P$ have been thoroughly and comprehensively explored.
At this point, we analyze two additional prevalent scenarios, specifically $L_m=-T$ and $L_m=\rho$. The purpose of considering $L_m=-T$ is therefore to investigate an alternative admissible representation of the matter sector and to examine how the choice of matter Lagrangian affects the resulting wormhole geometry and the corresponding energy conditions. Our objective is not to claim that $L_m=-T$ is the preferred matter Lagrangian, but rather to demonstrate that different admissible choices of $L_m$ lead to different physical predictions within the framework of $f(Q,T)$ gravity. For a perfect fluid one has $T=g^{\mu\nu}T_{\mu\nu}=-\rho+3p$ so $L_m=-T=\rho-3p$.
Thus $L_m=-T$ can be interpreted as a specific linear combination of $\rho$ and $p$. Nevertheless, this by itself does not establish its 'physicality'; to delve deeper, one must clarify why this particular combination is favored and how it relates to a coherent stress tensor and thermodynamic interpretation. This, however, falls outside the purview of this research.

In the case of $L_m=-T$, we derive the subsequent equations:
\begin{equation}\label{S1}
\rho=A_1\frac{b'}{r^2},
\end{equation}
\begin{equation}\label{S2}
p_r= \frac{A_2b+A_3rb'}{r^3},
\end{equation}
\begin{equation}\label{S3}
p_t=\frac{A_4b+A_5rb'}{r^3}.
\end{equation}
In another case of $L_m=\rho$, the following equations result:
\begin{equation}\label{S4}
\rho=A_6\frac{b'}{r^2},
\end{equation}
\begin{equation}\label{S5}
p_r= \frac{A_2b+A_7rb'}{r^3},
\end{equation}
\begin{equation}\label{S6}
p_t= \frac{A_4b+A_8rb'}{r^3}
\end{equation}
with
\begin{align}\label{SS}
A_1&=\left(\frac{\alpha}{1+\beta}\right)\frac{(1+2\beta)}{(1+3\beta)},\nonumber\\
A_2&=-\left(\frac{\alpha}{1+\beta}\right)=-\gamma(\alpha,\beta)=-2A_4,\nonumber\\
A_3&=\left(\frac{\alpha}{1+\beta}\right)\frac{\beta}{1+3\beta},\nonumber\\
A_5&=-\left(\frac{\alpha}{1+\beta}\right)\frac{(1+\beta)}{2(1+3\beta)},\nonumber\\
A_6&=\left(\frac{\alpha}{1+\beta}\right)\frac{1}{(1-2\beta)},\nonumber\\
A_7&=-\left(\frac{\alpha}{1+\beta}\right)\frac{2\beta}{(1-2\beta)},\nonumber\\
A_8&=\left(\frac{\alpha}{1+\beta}\right)\frac{(1+2\beta)}{2(1-2\beta)}.
\end{align}
It is evident that the overall structure of the field equation remains consistent across the three different selections for $L_m$; the sole variation occurs in the coefficient associated with the derivative of the shape function. In any case, various choices of $L_m$ result in different outcomes for the field equations. Nevertheless, as we have noted, in all three scenarios, the solutions that concurrently breach the conditions of $H$ and $H_1$ in GR may yield non-exotic solutions within the framework of $f(Q,T)$ theory.

\subsection{Solutions for $L_m =-T$}
In the following, we will investigate the solutions relevant to the cases $L_m=-T$ and $L_m=\rho$ and then compare with the results of Ref. \cite{fq-sara} for $L_m=-P$. Initially, we analyze the EoS $p_r=\omega \rho$ and subsequently, explore the issue concerning the variable EoS $p_r=\omega(r)\rho$.

\subsubsection{Solutions with Linear EoS}
\begin{figure}
\subfloat[$n_1>1$]{\includegraphics[width = 1.6in]{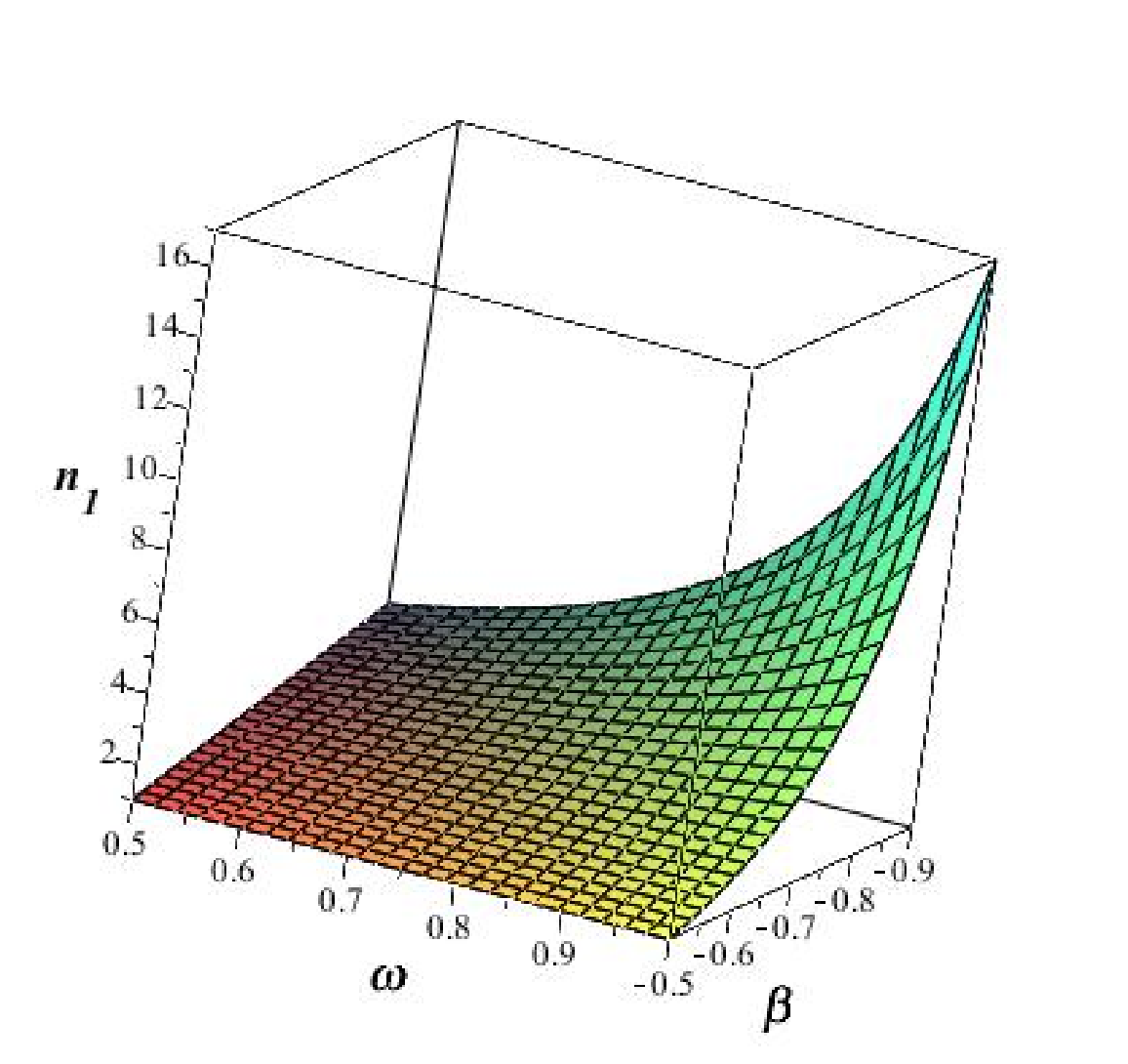}}\\
\subfloat[$n_1<1$]{\includegraphics[width = 1.6in]{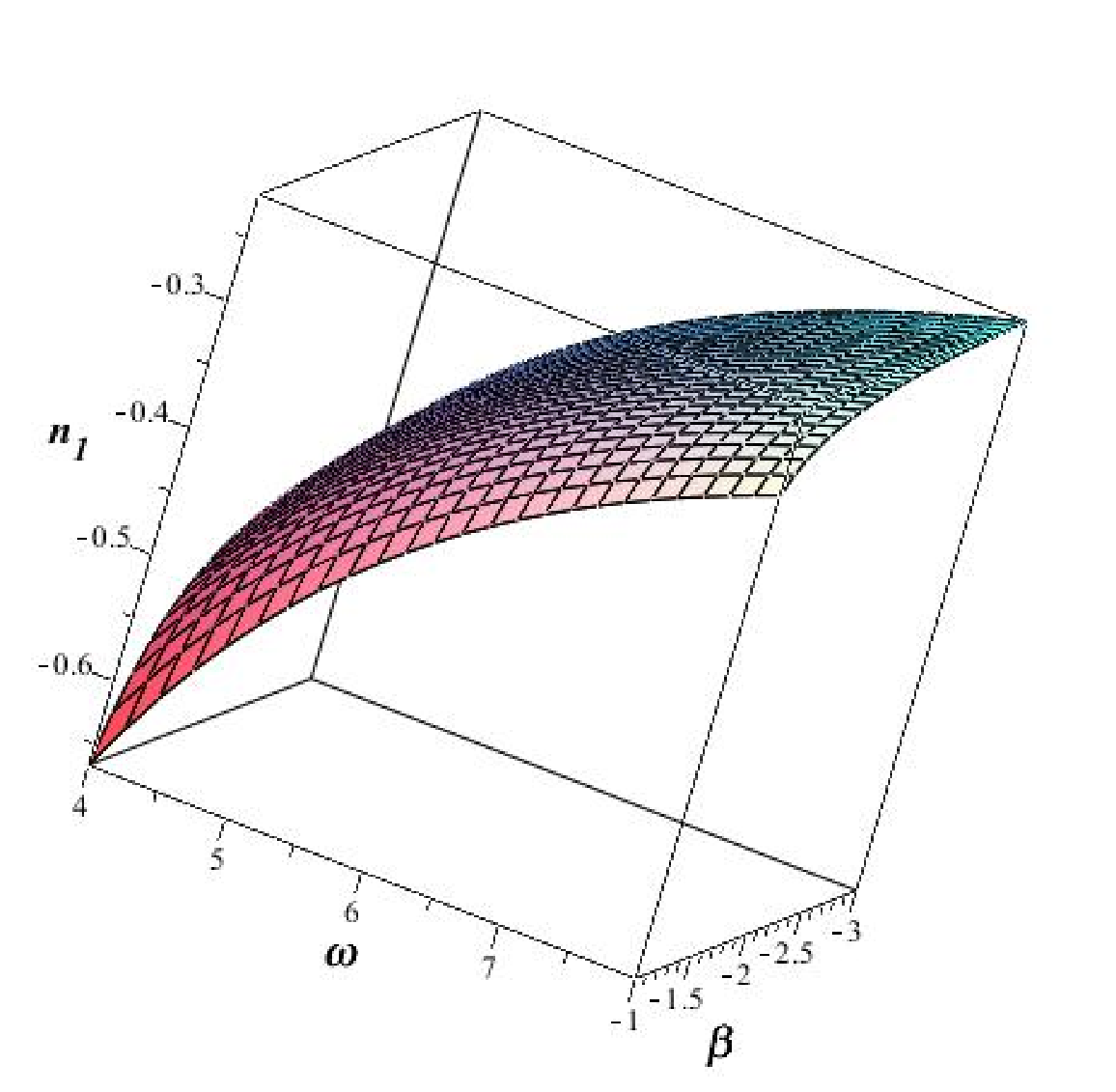}}\\
\subfloat[$n_1<1$]{\includegraphics[width = 1.6in]{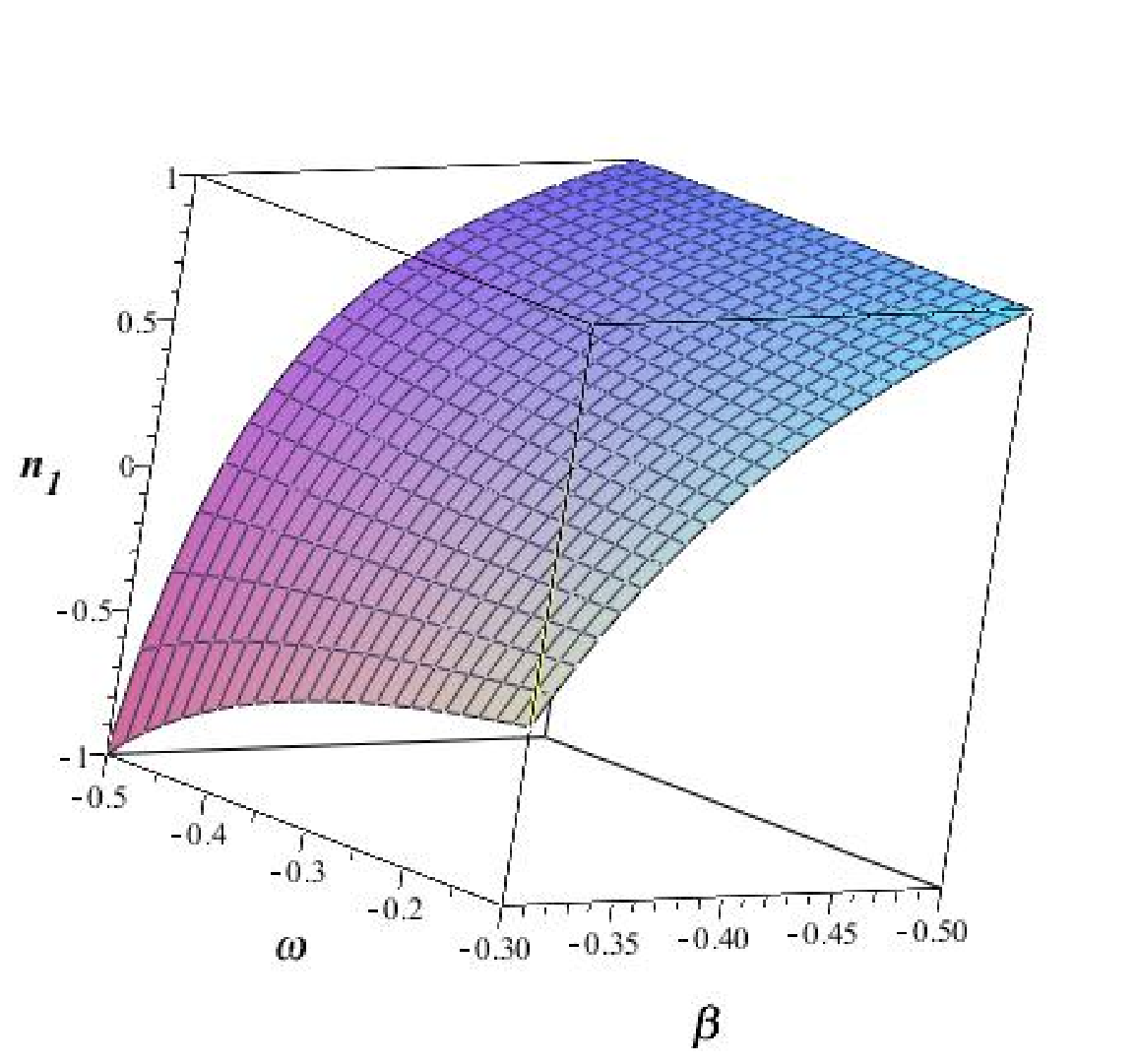}}\\
\subfloat[$n_1<1$]{\includegraphics[width = 1.6in]{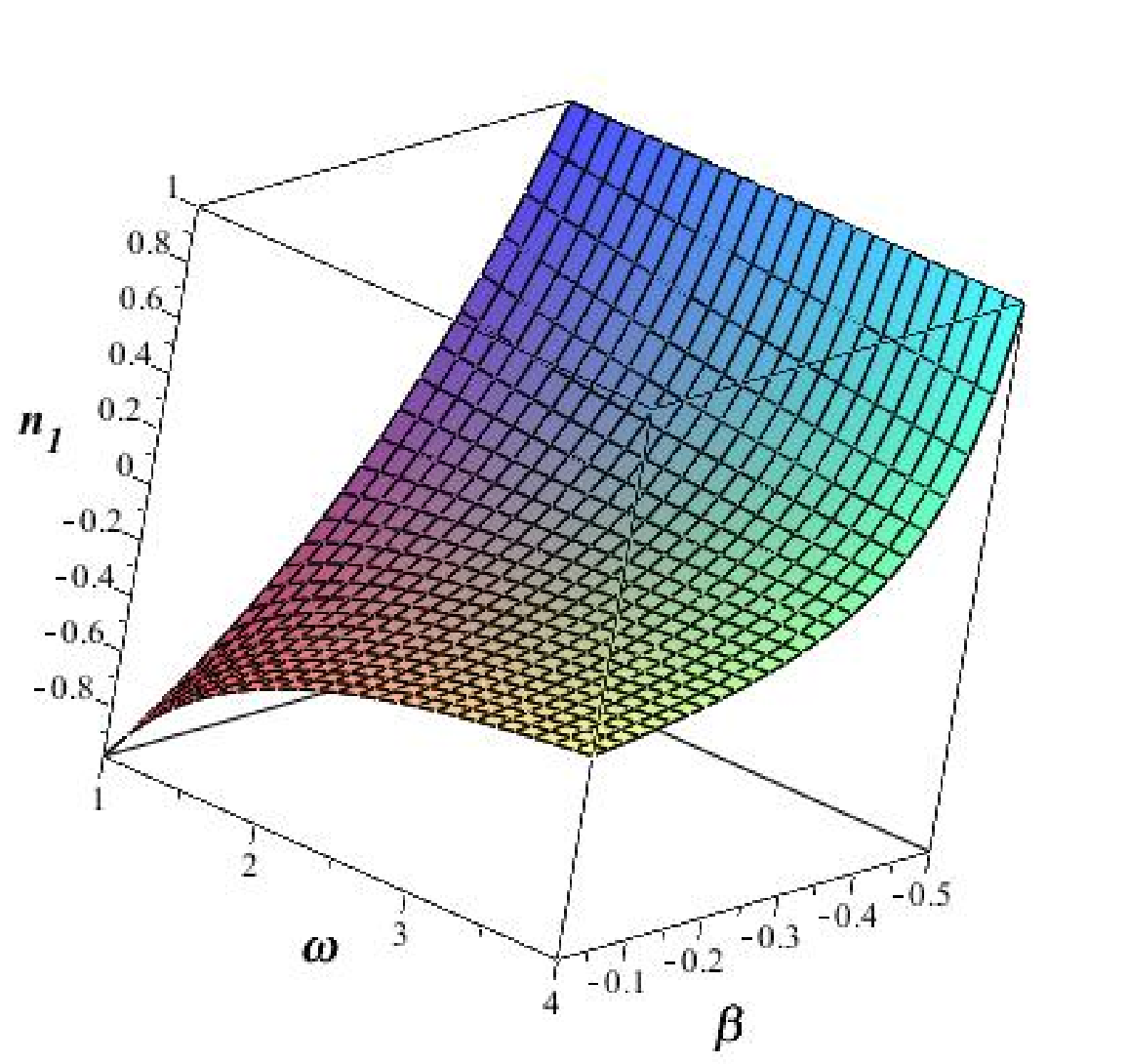}}\\
\subfloat[$n_1<1$]{\includegraphics[width = 1.6in]{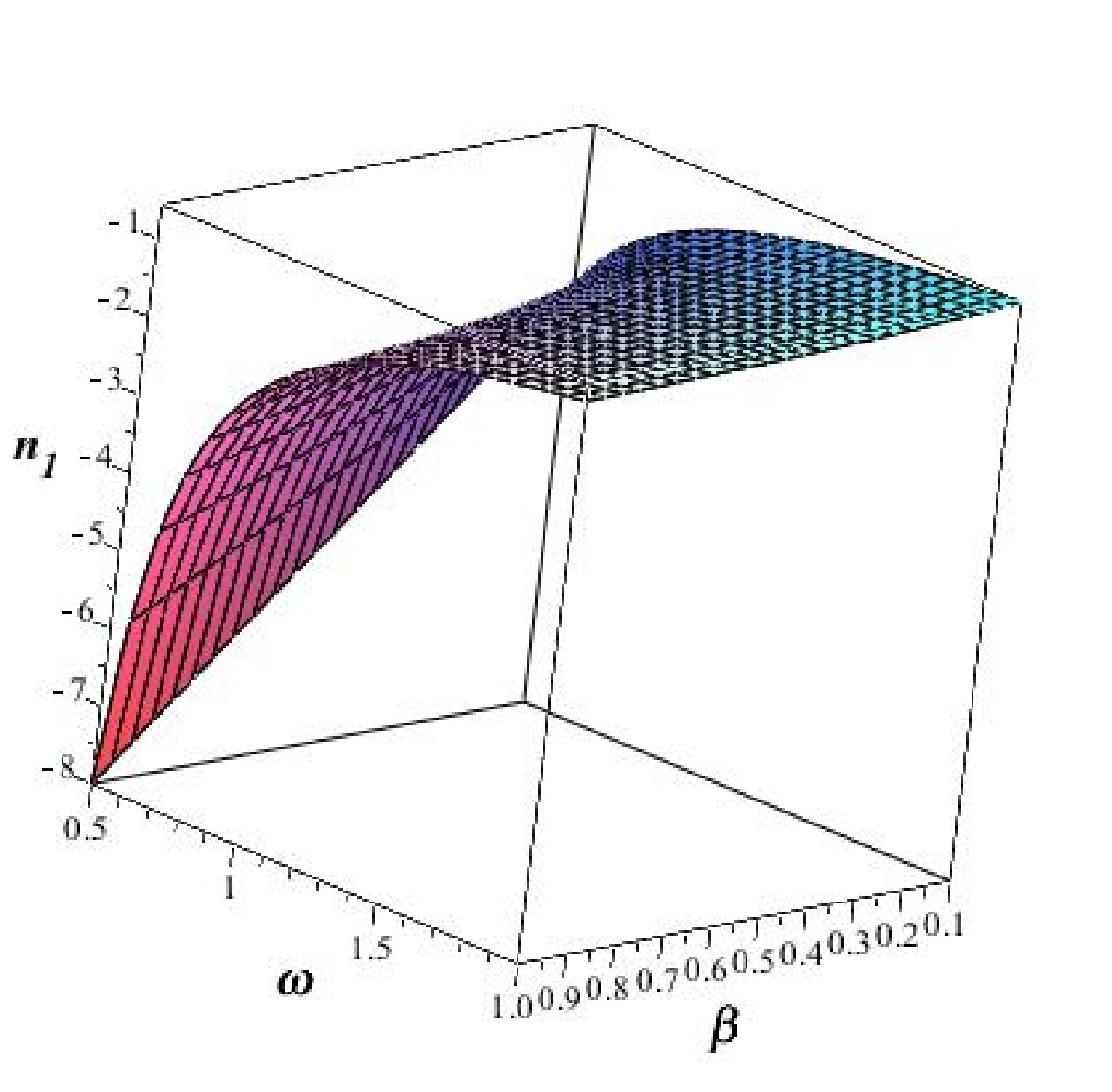}}\\
\caption{The graph depicts the correlation between $n_1(\omega,\beta)$ and the variables $\omega$ and $\beta$. It is clear that $n_1>1$ is valid within a certain range of $\omega$ and $\beta$ (a), while $n<1$ is acceptable in a different range (b, c, d, e).}s
\label{fig1}
\end{figure}

The linear EoS is pivotal in the study of wormhole physics, since it describes the material properties of the exotic matter that is necessary for sustaining a traversable geometry. Taking into account
\begin{equation}\label{17a}
p_r(r)=\omega\rho(r),
\end{equation}
and applying Eqs.(\ref{S1}) and (\ref{S2}) it can be obtained
\begin{equation}\label{18a}
b(r)=r^{n_1(\omega,\beta)},
\end{equation}
\begin{equation}\label{18b}
n_1(\omega,\beta)=\frac{1+3\beta}{\beta-\omega(1+2\beta)}.
\end{equation}
This function takes the shape of the well-known power-law function. If $n_1<1$, all the conditions required for a wormhole are satisfied. Investigating the ECs associated with this shape function reveals that to ensure a positive energy density, we must find intervals in which the following condition holds
\begin{equation}\label{19b}
n_1(\omega,\beta)A_1>0,
\end{equation}
It is obvious that
\begin{equation}\label{20a}
A_1=\gamma \frac{1+2\beta}{1+3\beta}.
\end{equation}
In light of the critical condition $\gamma<0$, the sign of the following term must be determined
\begin{equation}\label{21aa}
\frac{1+2\beta}{1+3\beta}.
\end{equation}
It can be illustrated that the expression (\ref{21aa}) holds a negative value in the interval
\begin{equation}\label{22a}
-\frac{1}{2}<\beta<-\frac{1}{3}
\end{equation}
and it is positive outside of it. It is understood that the condition $\omega>-1$ must be satisfied in order to present the solutions that comply with the WEC. The function of $n_1(\omega,\beta)$ has depicted in terms of $\omega$ and $\beta$ in Fig.(\ref{fig1}) in different intervals. It is apparent that $n_1(\omega,\beta)$ is acceptable within designated ranges of $\omega$ and $\beta$. Five choices have been chosen based on the quantifiable ranges of $\omega$ and $\beta$, as plotted in Fig.(\ref{fig1}).
Next, we have selected certain values for the free parameters $\alpha$, $\beta$, and $\omega$ then based on these selections, have investigated the ECs. In the primary case, we assign $\alpha=\frac{5}{2}\beta=2\omega=-1$ resulting in $b(r)=r^{(2/3)}$. The second option is $\alpha=-\frac{\beta}{2}=\frac{\omega}{6}=1$, which produces $b(r)=r^{(-5/16)}$. The third case consists of $\alpha=5\beta=-\frac{2}{3}\omega=-1$, yielding $b(r)=r^{(-4/11)}$. The next selection is $\alpha=-2\beta=-\frac{4}{5}\omega=-1$, leading to $b(r)=r^{(-5/4)}$. The last option is $\alpha=-5\beta=-\frac{4}{3}\omega=-1$, which gives $b(r)=r^{(-32/17)}$.
 These results are briefly illustrated in Table (\ref{Tab1}).
\begin{table*}[]
\begin{tabular}{|l|l|l|l|l|l|l|l|}
\hline
$b(r)$  & $\rho>0$ & $H>0$  & $H_1>0$ & $H_2>0$ & $H_3>0$ & $H_4>0$  & Satisfied ECs  \\ \hline
$r^{2/3}$  & \checkmark & \checkmark & $\times$ & \checkmark & $\times$ & $\times$ &  \\ \hline
 $r^{-5/16}$  & \checkmark & \checkmark & $\times$ & $\times$ & $\times$ & \checkmark &   \\ \hline
 $r^{(-4/11)}$ & \checkmark & \checkmark & $\times$ & $\times$ & $\times$ & $\times$ &  \\ \hline
$r^{-5/4}$ &  \checkmark & \checkmark & $\checkmark$ & $\times$ & $\checkmark$   & \checkmark  & $WEC$, $NEC$, $SEC$ \\ \hline
$r^{-32/17}$ &  \checkmark & \checkmark & \checkmark & $\checkmark$ & \checkmark   & \checkmark  & everyone \\ \hline
 \end{tabular}
\caption{The results of ECs for some individual $b(r)$.  }\label{Tab1}
\end{table*}

\subsubsection{Asymptotically linear EoS}\label{Sub2}
Although the linear EoS is the most common form typically considered for these equations, a broader range of solutions can be derived within the context of wormhole theory by employing a variable EoS. This strategy is predicated on the idea that asymptotically linear EoS should be considered a more comprehensive type of EoS rather than being limited to a strictly linear interpretation. It seems more physically reasonable to perceive the Universe as acting like a perfect fluid on a global scale instead of a local one. This approach has been employed to derive solutions in the domains of GR \cite{variable}, $f(R,T)$ gravity \cite{SR1}, $f(T,\mathcal{T})$ gravity \cite{foad3}, and $f(R,L_m)$ gravity \cite{SR2}. The identical algorithm is applicable for discovering asymptotically flat wormhole solutions in the realm of $f(Q, T)$ gravity. We will now examine a variable EoS as bellow
\begin{equation}\label{20b}
p_r(r)=\omega(r)\rho(r),
\end{equation}
with
\begin{equation}\label{21b}
\omega(r)=\omega_{\infty}+g(r).
\end{equation}
In this context, $\omega_{\infty}$  signifies the EoS parameter at considerable radial distances from the throat, and $g(r)$ is obligated to conform to
\begin{equation}\label{22b}
\lim_{r\longrightarrow \infty}g(r)=0.
\end{equation}
Condition $\rho>0$ along with $H\geq0$ implies
\begin{equation}\label{Cg1}
\omega(r)\geq-1.
\end{equation}
In light of the Eqs.(\ref{20b}), (\ref{21b}), (\ref{S1}) and (\ref{S2}) we will have
\begin{equation}\label{23a}
b(r)=C\exp\left(\int\frac{A_2dr}{r\left[(\omega_{\infty}+g(r))A_1-A_3)\right]} \right),
\end{equation}
which $C$ is a constant of integration. We will employ condition (\ref{2}) in order to identify this constant. By examining various form of $g(r)$, alternative solutions for $b(r)$ may be achieved, similar to reference procedure \cite{fq-sara}. Utilizing the reverse approach grants us the freedom to choose the shape function arbitrarily, followed by the identification of the EoS parameter.

Solutions which violate the radial and lateral NEC in the context of GR gravity are previously presented in \cite{SR1} in the context of $f(R,T)$ gravity. Some of the solutions presented in \cite{SR1} are analyzed in the context of $f(Q,T)$ gravity for $L_m=-P$ in \cite{fq-sara}. According to procedure presented in \cite{fq-sara}, we analyze the same known shape functions for $L_m=-T$ in this Section. It should be noted that the energy density $\rho$ is required to be positive and according to Ref. \cite{SR1} the energy density is as follows
\begin{equation}
\rho(r)=\frac{1}{1+2\lambda}\frac{b'}{r^2}.
\end{equation}
Taking into account that the term $1+2\lambda$ is negative \cite{SR1}, the ensuing condition must be fulfilled
\begin{equation}
\frac{b'}{r^2}<0,
\end{equation}
so considering relations in (\ref{S1}) and (\ref{SS}), it concludes
\begin{equation}\label{24a}
\frac{A_1}{\gamma}=\frac{1+2\beta}{1+3\beta}>0,
\end{equation}
in the preceding subsection, it was indicated that this condition is valid in the area beyond the specified range, $-\frac{1}{2}<\beta<-\frac{1}{3}$.
To exemplify, we take two shape functions, as previously discussed in \cite{SR1} and \cite{fq-sara}, and assess their physical attributes within the framework of $f(Q, T)$ gravity.
The first case is as follows
\begin{equation}\label{25b}
b(r)=(\omega+D)^{\frac{1}{\omega}}(\omega r+D)^{-\frac{1}{\omega}}.
\end{equation}
This shape function contravenes both radial and lateral NEC within the framework of GR, making it a potential candidate to fulfill the NEC in the domain of $f(Q,T)$ gravity.

\begin{figure}
\centering
  \includegraphics[width=3 in]{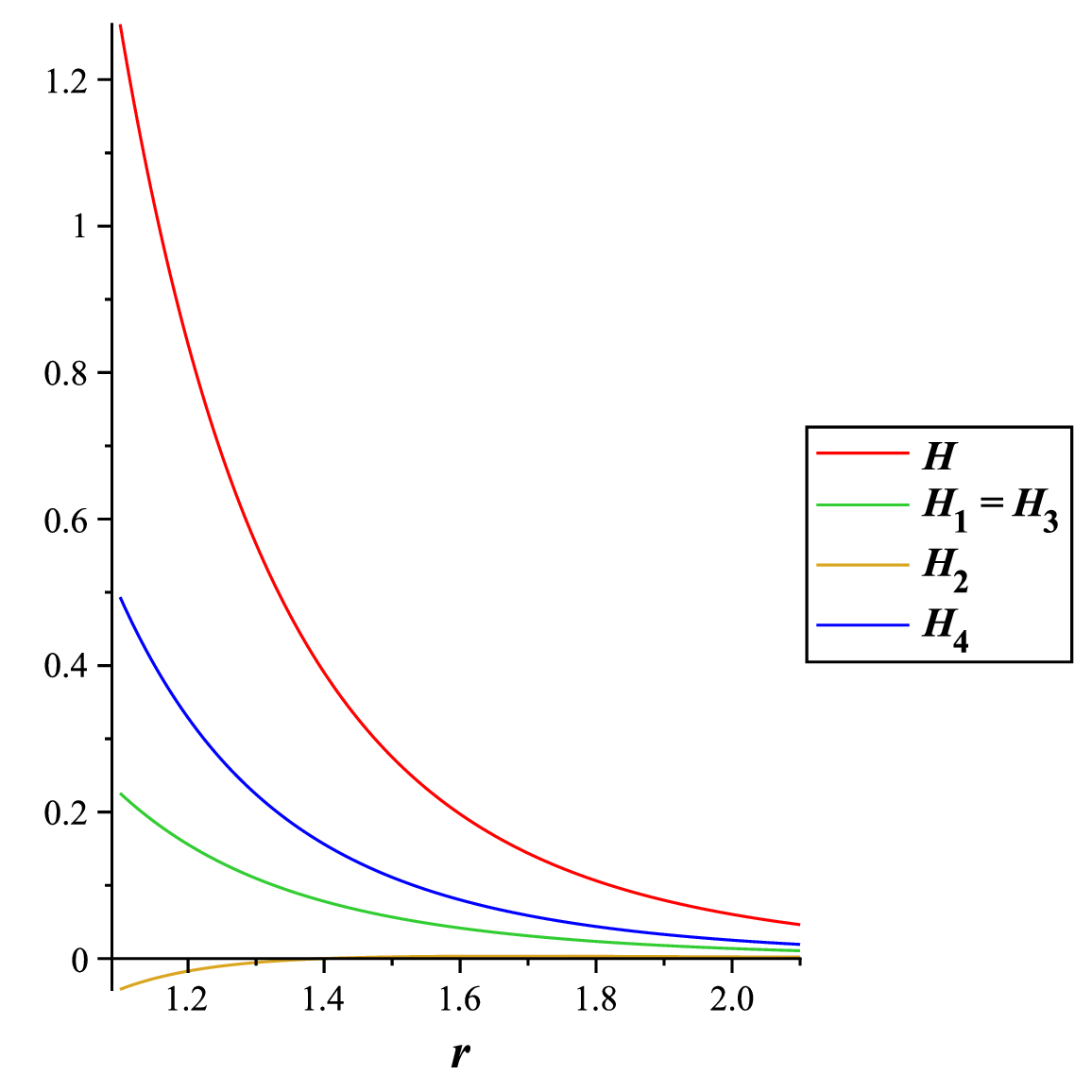}
\caption{The plot depicts $H(r)$ (red), $H_1(r)=H_3(r)$ (green),  $H_2(r)$ (yellow), and $H_4(r)$ (blue) against $r$ for $b(r)=\left(\frac{2}{r+1}\right)^4$ which shows that all the ECs except the DEC are satisfied. See the text for details.}
 \label{fig2}
\end{figure}

\begin{figure}
\subfloat[]{\includegraphics[width = 2in]{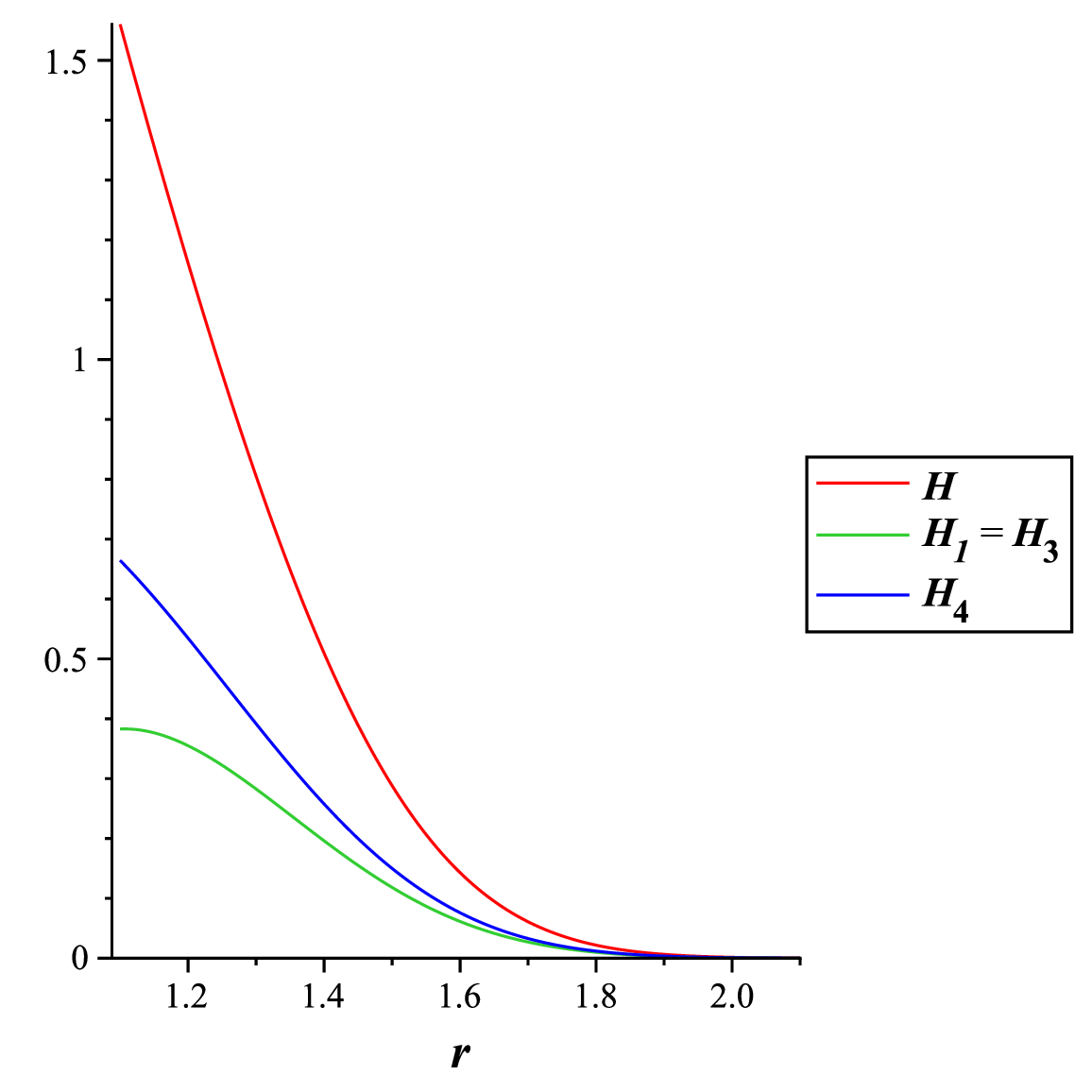}}\\
\subfloat[]{\includegraphics[width = 2in]{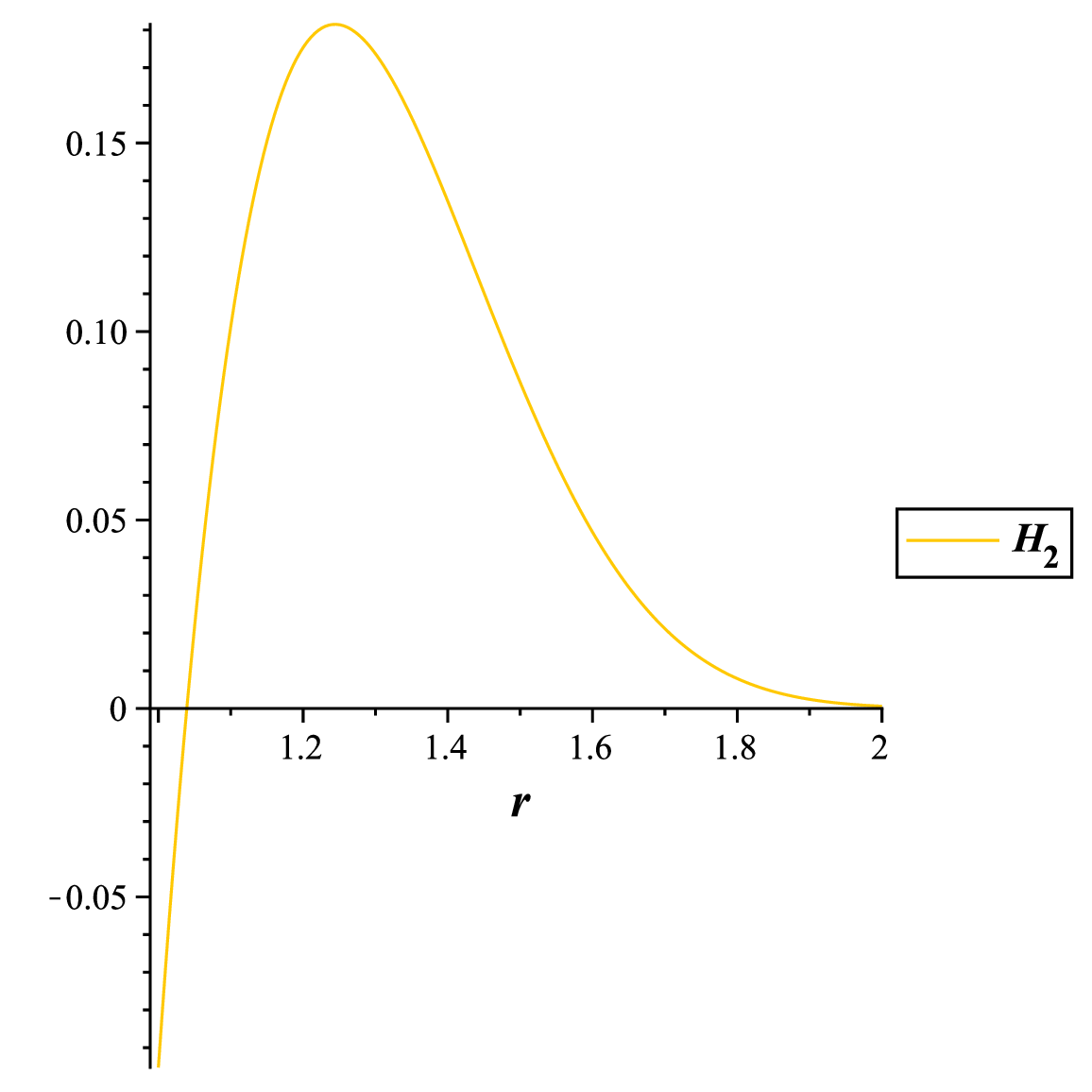}}\\
\caption{The plot depicts $H(r)$ (red), $H_1(r)=H_3(r)$ (green),  $H_2(r)$ (yellow), and $H_4(r)$ (blue) against $r$ for $b(r)=\left(\frac{2}{r+1}\right)^4$ which shows that all the ECs except the DEC are satisfied. See the text for details.}s
\label{fig3}
\end{figure}

\begin{figure}
\centering
  \includegraphics[width=3 in]{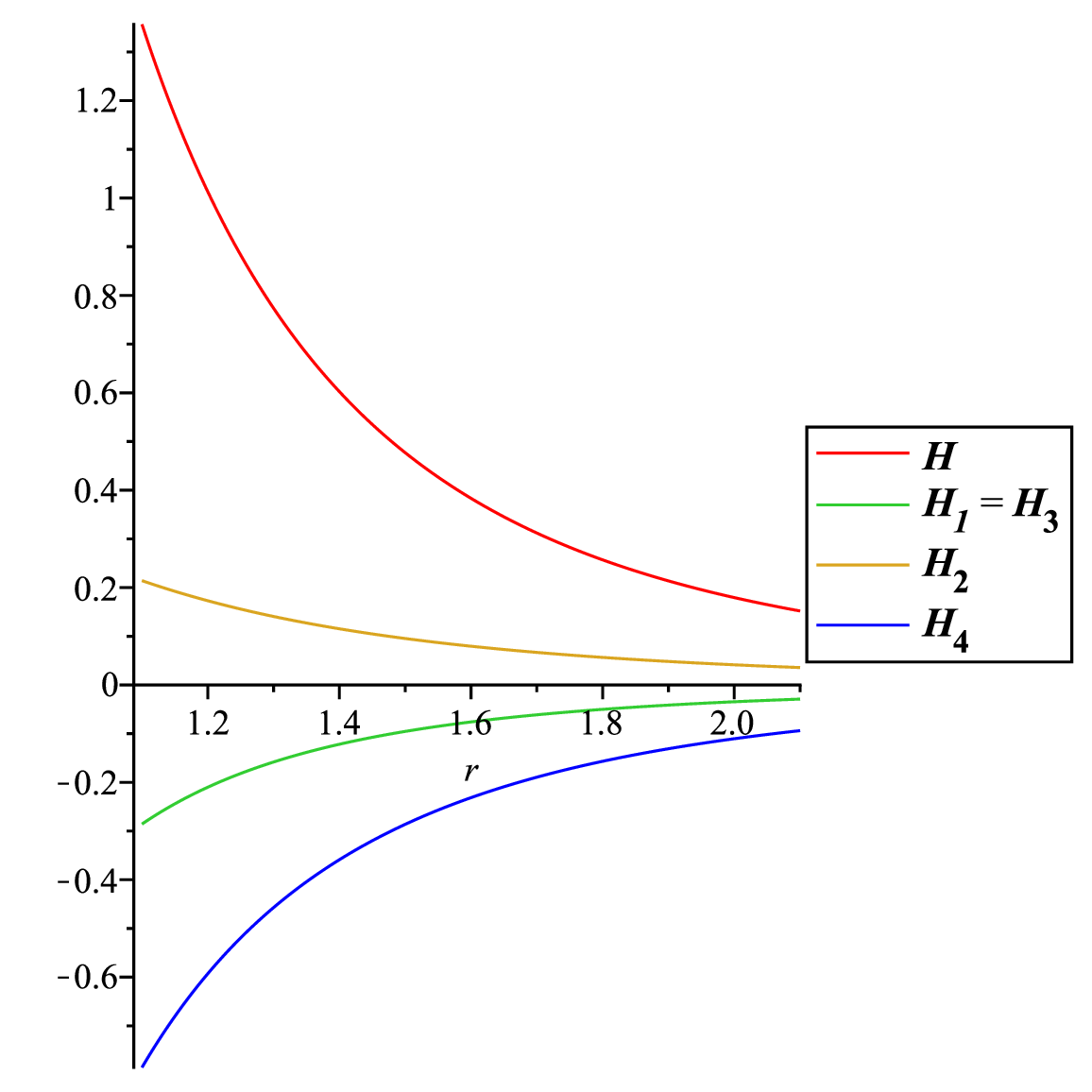}
\caption{The plot depicts the functions $ H(r)$(red), $H_1(r)=H_3(r)$(green), $H_2(r)$(yellow), and $H_4(r)$(blue) plotted against the radial coordinate for the shape function $b(r)=\sqrt{ \frac{5}{4r+1}}$. It is evident that all ECs are violated. See the text for details.}
 \label{fig4}
\end{figure}

\begin{figure}
\centering
  \includegraphics[width=3 in]{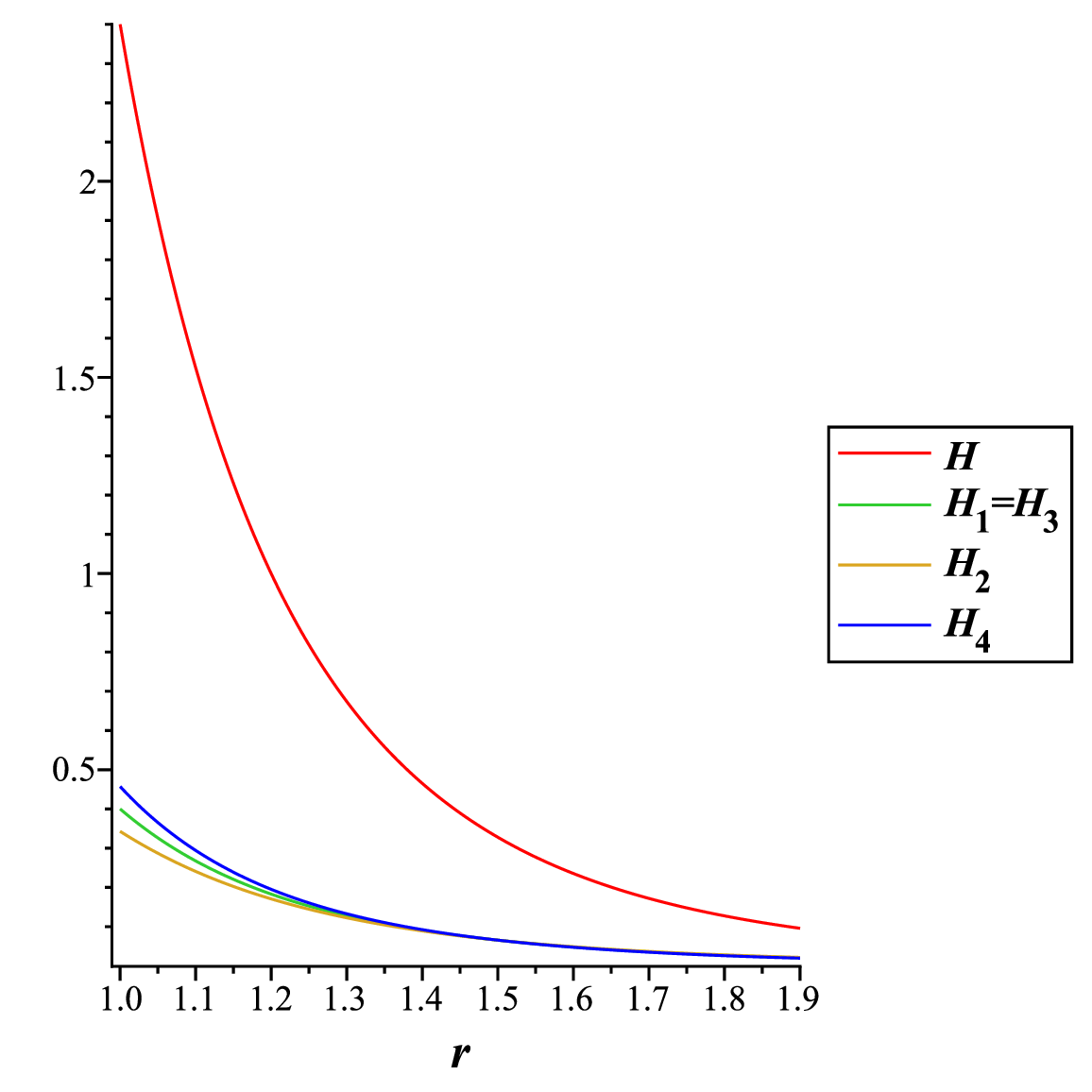}
\caption{The graph depicts the functions $ H(r)$(red), $H_1(r)=H_3(r)$(green), $H_2(r)$(yellow), and $H_4(r)$(blue) plotted against the radial coordinate for the shape function $b(r)= \frac{343\sqrt{7}}{(4r+3)^{7/2}}$. It is evident that all ECs are respected. See the text for details.}
 \label{fig5}
\end{figure}

Now, we assign $\alpha=-\beta=-8D=-8\omega=-2$, leading to
\begin{equation}\label{26b}
b(r)=\left(\frac{2}{r+1}\right)^4,
\end{equation}
which has been explored in \cite{fq-sara}. But the energy density is given by
\begin{equation}\label{27b}
\rho(r)=\frac{640}{21r^2(r+1)^5},
\end{equation}
where is different from the results achieved in \cite{SR1} and \cite{fq-sara}. The energy density is positive; therefore, we can plot $H, H_1, H_2, H_3$, and $H_4$ in terms of radial coordinate in Fig.(\ref{fig2}). This representation denotes that all ECs, excluding the DEC, are recognized in the context of $f(Q,T)$ for the shape function (\ref{26b}). Currently, we are able to identify the EoS parameter in the following manner
\begin{equation}\label{28b}
\omega_{\infty}+g(r)=\frac{p(r)}{\rho(r)}=\frac{15r+7}{20r}.
\end{equation}
It can be observed that $\omega_{\infty}=\frac{3}{4}$ and $g(r)=\frac{7}{20r}$, which is not the same as the mentioned result in \cite{SR1} and \cite{fq-sara}.
The next example is
\begin{equation}\label{29b}
b(r)= \exp(\frac{1}{nD}-\frac{r^n}{nD}).
\end{equation}
This shape function fulfills all the ECs within the framework of $f(R, T)$ gravity \cite{SR1} and also adheres to all of the ECs, with the exception of the DEC, in the context of $f(Q,T)$ for $L_m=-P$ \cite{fq-sara}. We now will assess this shape function in the same manner as demonstrated in the earlier example. Within the scope of $f (Q, T)$ gravity, we define $\alpha=-\beta=-4D=-\frac{n}{2}=-2$, which results in
\begin{equation}\label{30b}
b(r)=\exp(\frac{1-r^4}{2}),
\end{equation}
by applying (\ref{30b}) in (\ref{S1}) we have
\begin{equation}\label{31b}
\rho(r)=\frac{20}{21}r\exp(\frac{1-r^4}{2}).
\end{equation}
Since the energy density is positive, we have represented $H, H1, H2, H3$, and $H4$ in relation to the radial coordinate as shown in Fig. (\ref{fig3}). This representation signifies that all ECs, excluding the DEC, are met within the context of $f(Q, T)$ for the shape function (\ref{30b}). It is simple to prove that the shape function (\ref{29b}) provides
\begin{equation}
\omega(r)=\frac{2}{5}+\frac{7}{10r^4}.
\end{equation}
This $\omega(r)$ differs from the findings presented in \cite{SR1} and \cite{fq-sara}. In the next step, we follow the second method. We shall examine the solutions with a recognized EoS, as our initial choice, we will consider
\begin{equation}\label{32b}
g(r)=\frac{D}{r},
\end{equation}
which leads to
\begin{equation}\label{33b}
b(r)=\left(B_1+D_1\right)^{-n_1(\omega_{\infty}, \beta)}\left(B_1r^n+D_1\right)^{n_1(\omega_{\infty}, \beta)},
\end{equation}
where
\begin{equation}\label{A1}
B_1=(1+2\beta)\omega_{\infty}-\beta,
\end{equation}
\begin{equation}\label{A2}
D_1=(1+2\beta) D,
\end{equation}
\begin{equation}\label{n1}
n_1(\omega_{\infty}, \beta)=\frac{1+3\beta}{n(\beta-\omega_{\infty}(1+2\beta))}.
\end{equation}
It is evident that the general structure of $b(r)$ resembles what was discovered in \cite{fq-sara}, yet the parameters $B_1$, $D_1$, and $n_1$ are entirely distinct.
For large distances, the condition (\ref{Cg1}) transform to
\begin{equation}
\omega_{\infty}\geq-1.
\end{equation}
As the first case, we set $\alpha=4\beta=-2\omega_{\infty}=-4D=-1$ which leads to
\begin{equation}\label{34b}
b(r)=\sqrt{ \frac{5}{4r+1}}
\end{equation}
and consequently
\begin{equation}\label{35b}
\rho(r)=\frac{16\sqrt{5}}{3(4r+1)^{3/2}r^2}.
\end{equation}
We have plotted $H, H1, H2, H3$, and $H4$ as a function of radial coordinate in Fig. (\ref{fig4}). This figure demonstrates that none of the ECs are respected. The same choices for $\alpha, \beta, \omega_{\infty},$ and  $D$ leads to a different shape function for $l_m=-P$ while all the ECs are satisfied \cite{fq-sara}. By using $\alpha=-4\beta=-2\omega_{\infty}=-4D=-1$, one can obtain
\begin{equation}\label{36b}
b(r)= \frac{343\sqrt{7}}{(4r+3)^{7/2}}.
\end{equation}
The corresponding energy density is
 \begin{equation}\label{37b}
\rho(r)=\frac{16464\sqrt{7}}{5(4r+3)^{9/2}r^2}.
\end{equation}
Figure (\ref{fig5}) demonstrate the behaviour of the ECs with respect to radial coordinate. It is clear that all of the ECs are respected for the shape function (\ref{36b}).

\subsection{Solutions for $L_m=\rho$}
Currently, we examine the wormhole solutions based on the assumption $L_m= \rho$ within the framework of $f(Q,T)$ gravity. The procedure remains consistent with the previous section where we analyzed the solutions for $l_m=-T$.

\subsubsection{Solutions with Linear EoS}
For a linear EoS by using Eqs. (\ref{S4}) and (\ref{S5}) in EoS (\ref{17a}), we will have
\begin{equation}\label{1c}
b(r)=r^{n_2(\omega,\beta)},
\end{equation}
in which
\begin{equation}\label{2c}
n_2(\omega,\beta)=\frac{2\beta-1}{\omega+2\beta}.
\end{equation}
It is obvious that it fulfills all the conditions for wormhole theory provided that $n_2<1$ holds true. To guarantee a positive energy density, it is essential that the bellow condition must be satisfied
\begin{equation}\label{3c}
n_2(\omega,\beta)A_6>0.
\end{equation}
It can be observed that
\begin{equation}\label{4c}
A_6=\frac{\gamma}{1-2\beta}.
\end{equation}
Since $\gamma$ is negative, the sign of the ensuing expression must be identified
\begin{equation}\label{5c}
\frac{1}{1-2\beta}.
\end{equation}
It can be evidenced that the expression (\ref{5c}) possesses a negative value for
\begin{equation}\label{6c}
\beta>\frac{1}{2},
\end{equation}
and it is positive for $\beta<\frac{1}{2}$. Same as $L_m=-T$, $\omega>-1$ must be preserved. The $n_2(\omega,\beta)$ function can be plotted with respect to $\omega$ and $\beta$. Following this, we have identified particular values for the free parameters $\alpha$, $\beta$, and $\omega$, and based on these selections, we have analyzed the ECs. In the first instance, we designate  $\alpha=\frac{3}{5}\beta=\frac{1}{2}\omega=-1$ which leads to $b(r)=r^{(13/16)}$.
The next instance is  $\alpha=-4\beta=-\frac{1}{2}\omega=-1$, yielding $b(r)=r^{(-1/5)}$. The concluding option is $\alpha=-2\beta=4\omega=-2$, producing $b(r)=r^{(2/3)}$. These findings are briefly summarized in Table (\ref{Tab2}).

\begin{table*}[]
\begin{tabular}{|l|l|l|l|l|l|l|l|}
\hline
$b(r)$  & $\rho>0$ & $H>0$  & $H_1>0$ & $H_2>0$ & $H_3>0$ & $H_4>0$  & Satisfied ECs  \\ \hline
 $r^{13/16}$  & \checkmark & $\times$ & \checkmark & $\times$ & $\times$ & \checkmark &   \\ \hline
 $r^{-1/5}$  & \checkmark & \checkmark & \checkmark & $\times$ & \checkmark & \checkmark & $WEC$, $NEC$, $SEC$  \\ \hline
$r^{2/3}$ &  \checkmark & \checkmark & \checkmark & $\checkmark$ & \checkmark   & \checkmark  & everyone \\ \hline
 \end{tabular}
\caption{The results of ECs for some $b(r)$ with $L_m=\rho$ while liner EoS is assumed. }\label{Tab2}
\end{table*}

\subsubsection{Asymptotically linear EoS}
Now, we will investigate a variable EoS as indicated in Eq.(\ref{20b}), and reiterate the same strategy as presented in section \ref{Sub2}.
Through the examination of various forms of $g(r)$, alternative solutions for $b(r)$ can be realized. Considering that the energy density must be positive, the coefficient of equation (\ref{S4}) is obligated to meet the following equation which is the equivalent relation to Eq.(\ref{24a}) :
\begin{equation}\label{8c}
\frac{A_6}{\gamma}=\frac{1}{1-2\beta}>0
\end{equation}
This condition is satisfied for
\begin{equation}\label{7c}
\beta<\frac{1}{2}.
\end{equation}
Currently, we are looking into two special shape functions and addressing their physical properties within the context of $f(Q,T)$ theory.

\begin{figure}
\centering
  \includegraphics[width=3 in]{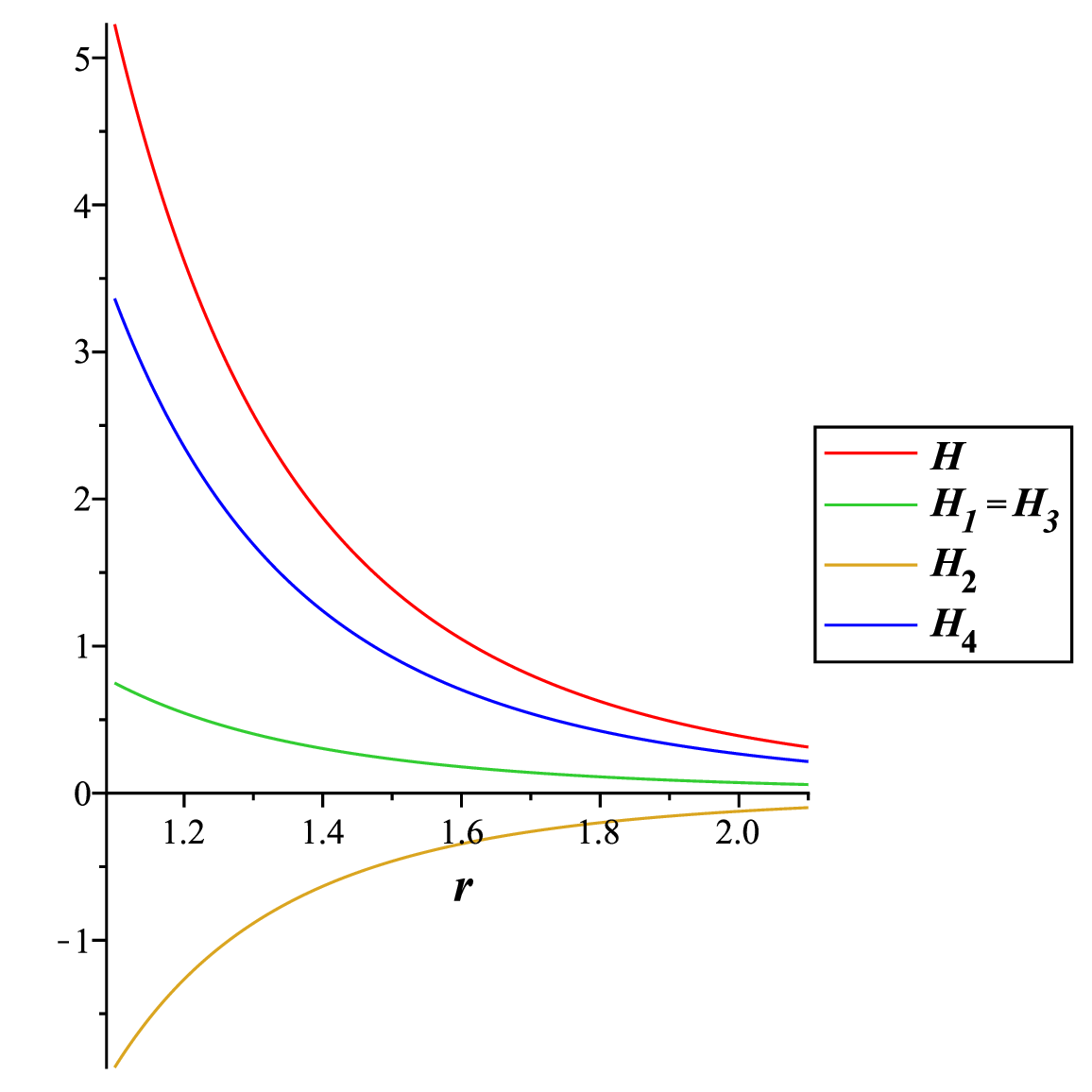}
\caption{The graph depicts the functions $ H(r)$(red), $H_1(r)=H_3(r)$(green), $H_2(r)$(yellow), and $H_4(r)$(blue) plotted against the radial coordinate for the shape function $b(r)=\left(\frac{3}{2r+1}\right)^2$. It is evident that all ECs except the DEC are respected. See the text for details.}
 \label{fig6}
\end{figure}

We will initiate with the given shape function in Eq. (\ref{25b}). By choosing $\alpha=-\beta=-8D=-8\omega=-2$, results in
\begin{equation*}
b(r)=\left(\frac{2}{r+1}\right)^4,
\end{equation*}
in this case the condition $\beta<\frac{1}{2}$ is not valid. It is apparent that the energy density is negative and all the ECs are violated. This outcome is entirely distinct from what we have previously discovered for $L_m=-P$ and $L_m=-T$ with the same shape function. As another choice, we set $\alpha=5\beta=-8D=-4\omega=-2$ which gives
 \begin{equation}\label{fc}
b(r)=\left(\frac{3}{2r+1}\right)^2,
\end{equation}
and
\begin{equation}\label{fcc}
\rho(r)=\frac{200}{3(2r+1)^3r^2}.
\end{equation}
The $H, H_1, H_2, H_3$ and $H_4$ is plotted as a function of $r$ in Fig. (\ref{fig6}) which indicates that all of the ECs except DEC are satisfied for the shape function (\ref{fc}).
We are now going to examine the shape function (\ref{29b}) in a manner similar to that illustrated in the earlier example. We specify $\alpha=-\beta=-4D=-\frac{n}{2}=-2$, resulting in (\ref{30b}). Using Eq.(\ref{S4}) we obtain
\begin{equation}\label{10c}
\rho(r)=-\frac{4}{9}r\exp(\frac{1-r^4}{2}).
\end{equation}
It is evident that $\rho$ is negative and  all of the ECs continue to be unsatisfied, which contrasts with the scenario for $L_m=-P$ and $L_m=-T$. It is easy to show that the case,  $\alpha=-\beta=-4D=-\frac{n}{2}=-2$  provides
 \begin{equation}\label{fc1}
b(r)= \exp(1-r^4),
\end{equation}
and
\begin{equation}\label{fc2}
\rho(r)=\frac{200}{27}r\exp(1-r^4).
\end{equation}
The analyses of the shape function (\ref{fc1}) shows that this shape function admits all of the ECs except DEC.

In the next step, we will consider the solutions characterized by the specific EoS. As an first case, We assume $g(r)$ in the form of (\ref{32b}). In this case, we reach
\begin{equation}\label{11c}
b(r)=\left(B_2+D_2\right)^{-n_2(\omega_{\infty},\beta)}\left(B_2r+D_2\right)^{n_2(\omega_{\infty}, \beta)}
\end{equation}
with
\begin{equation}\label{12c}
B_2=2\beta+\omega_{\infty},
\end{equation}
\begin{equation}\label{13c}
n_2(\omega_{\infty}, \beta)=\frac{2\beta-1}{\omega_{\infty}+2\beta}.
\end{equation}
One can see that the general structure is the same as (\ref{33b}) but the coefficients are different from (\ref{A1}) and (\ref{A2}). The case $\alpha=4\beta=-2\omega_{\infty}=-4D=-1$ leads to
\begin{equation}\label{14c}
b(r)=\exp(6-6r)
\end{equation}
and
\begin{equation}\label{15c}
\rho(r)=\frac{16\exp(6-6r)}{3r^2}.
\end{equation}
Our analyses show that the shape function (\ref{14c}) violates the ECs. In the seconde case, $\alpha=-4\beta=-2\omega_{\infty}=-4D=-1$ gives
\begin{equation}\label{16c}
b(r)=\sqrt{ \frac{5}{4r+1}}
\end{equation}
and
\begin{equation}\label{17c}
\rho(r)=\frac{16\sqrt{5}}{5(4r+1)^{3/2}r^2}.
\end{equation}
It is notable that the shape function (\ref{16c})  is identical to the shape function (\ref{34b}) but the energy density is different from (\ref{35b}). The shape function (\ref{16c}) satisfies all of the ECs.


\section{Wormhole solutions for non-linear $f(Q,T)$}\label{sec4a}
In this research, we employ the linear representation of the function $f(Q,T)$. Asymptotically flat wormhole solutions can also be explored by considering nonlinear forms of the $f(Q,T)$ function.
The transition from a linear model of the form (\ref{99a}) to a non-linear functional framework significantly elevates the analytical and computational complexity of the resulting field equations. While the linear model maintains a structure that allows for a tractable exploration of the underlying matter-geometry coupling, non-linear dependencies—such as power-law or exponential terms—introduce higher-order derivatives of the non-metricity scalar $Q$ and $T$. These terms typically generate highly non-linear, coupled partial differential equations that are no longer amenable to standard exact solutions. Furthermore, the non-linear coupling terms often exacerbate the breakdown of the standard energy-momentum conservation law, necessitating the calculation of complex, non-local source terms that obscure the physical interpretation of the gravitational interaction. Consequently, the linear model provides a necessary and essential ‘first-order’ laboratory for isolating the specific effects of matter-geometry interaction, whereas non-linear models often require numerical approximation methods to discern the competing influence of geometric non-linearity from the underlying matter Lagrangian contribution. As an example, we assume
\begin{equation}\label{199a}
f(Q, T)=\alpha  Q+\beta T+\gamma T^2.
\end{equation}
By utilizing metric (\ref{1}) in conjunction with equations (\ref{6a})–(\ref{a5}), the resulting field equations are given by:
\begin{eqnarray}\label{18}
 \frac{2 (r-b)}{(2 r-b) f_Q}\Bigg[\rho+f_T (P+\rho )\nonumber \\
 -\frac{(r-b)}{ r^3} \Bigg(b f_Q \left(\frac{ r \phi' +1}{r-b}-\frac{2 r-b}{2 (r-b)^2}\right) \nonumber \\
+\frac{b r f_{\text{QQ}} Q'}{r-b} +\frac{f r^3}{2 (r-b)}\Bigg)\Bigg]=\frac{b'}{ r^2},
 \end{eqnarray}
\begin{eqnarray}\label{19}
\frac{2 b}{f r^3}\Bigg[p_r +\frac{(r-b)}{2 r^3} \Bigg( f_Q \Bigg(\frac{b \left(\frac{r b'-b}{r-b}+2 r \phi' +2\right)}{r-b}-4 r \phi'\Bigg)\nonumber \\
+\frac{2 b r f_{\text{QQ}} Q'}{r-b}\Bigg)
+\frac{f r^3 (r-b)\phi'}{ b r^2}-f_T \left(P-p_r\right)\Bigg]
\nonumber \\ =\Bigg[2\left(1-\frac{b}{r}\right)\frac{\phi'}{r}-\frac{b}{r^3}\Bigg],
 \end{eqnarray}
\begin{eqnarray}\label{20}
 \frac{1}{f_Q \left(\frac{r}{r-b}+r \phi' \right)}\Bigg[p_t +\frac{(r-b)}{4 r^2}\Bigg(f_Q \Bigg(\frac{4 (2 b-r) \phi'}{r-b}-4 r (\phi')^2 \nonumber \\
 -4 r \phi''\Bigg)+\frac{2 f r^2}{r-b}-4 r f_{\text{QQ}} Q' \phi' -f_T \left(P-p_t\right)  \nonumber \\
 +\frac{(r-b)}{ r}(\phi'' +{\phi'}^2-\frac{(rb'-b)\phi'}{2r(r-b)}+\frac{\phi'}{r})f_Q (\frac{r}{r-b}+r \phi'    )\Bigg]\nonumber \\
    =\left(1-\frac{b}{r}\right)\left[\phi'' +{\phi'}^2-\frac{(rb'-b)\phi'}{2r(r-b)}-\frac{rb'-b}{2r^2 (r-b)}+\frac{\phi'}{r}\right]. \end{eqnarray}
Setting $\phi=0$ and performing straightforward algebraic simplifications, we obtain:
\begin{eqnarray}\label{F19aa}
\rho = \frac{(r-b)}{2 r^3}  \Bigg[f_Q \left(\frac{(2 r-b) (r b'-b)}{(r-b)^2}+\frac{2b}{r-b}\right)
 \nonumber \\
+\frac{2 b r f_{\text{QQ}} Q'}{r-b}+\frac{f r^3}{r-b}-\frac{2r^3 f_T (-L_m+\rho )}{(r-b)} \Bigg],
\end{eqnarray}
\begin{eqnarray}\label{F11cc}
 p_r = -\frac{(r-b)}{2 r^3} \Bigg[f_Q \frac{b }{r-b}\left(\frac{r b'-b}{r-b}+2\right) \nonumber \\
 +\frac{2 b r f_{\text{QQ}} Q'}{r-b}+\frac{f r^3}{r-b}-\frac{2r^3 f_T (-L_m-p_r)}{(r-b)}\Bigg],
 \end{eqnarray}
\begin{eqnarray}\label{F11d}
 p_t = -\frac{(r-b)}{4 r^2} \Bigg[2f_Q \left(\frac{r b'-b}{(r-b)^2}\right)+\frac{2 f r^2}{r-b}\nonumber \\
 -\frac{4r^2 f_T (-L_m-p_t)}{(r-b)}\Bigg].
 \end{eqnarray}
Finally, substituting Eq.(\ref{199a}) into Eqs. (\ref{F19aa})–(\ref{F11d}), we arrive at:
\begin{eqnarray}\label{F3e}
  \rho =\frac{1}{2 r^3}[\alpha \left(\frac{(2 r-b) \left(r b'-b\right)}{(r-b)}+2b\right) \nonumber \\
 +f r^3+2r^3 (\beta+2\gamma T) (L_m-\rho )\Bigg],
\end{eqnarray}
\begin{eqnarray}\label{F3f}
 p=-\frac{1}{2 r^3} \Bigg[\alpha b\left(\frac{r b'-b}{r-b}+2\right) \nonumber \\
 +f r^3+2r^3 (\beta+2\gamma T) \left(L_m+p_r\right)\Bigg],
\end{eqnarray}
\begin{eqnarray}\label{F3g}
  p_t=-\frac{1}{2 r^2} \Bigg[\alpha \left(\frac{\left(r b'-b\right) }{r-b }\right)+f r^2\nonumber \\
  -2r^2 (\beta+2\gamma T) \left(-L_m-p_t\right)\Bigg].
\end{eqnarray}
By considering a linear EoS for pressure,
\begin{equation}\label{9a}
p_r (r)=\omega_r\rho
\end{equation}
and
\begin{equation}\label{9b}
p_t (r)=\omega_t\rho.
\end{equation}
where $\omega_r$ and $\omega_t$ are the constant EoS parameters.  For $L_m=-P$, substituting Eqs. (\ref{9a}) and (\ref{9b}) into the field equations (\ref{F3e}–\ref{F3g}) yields
 \begin{eqnarray}
\frac{b^{\prime}(r)}{r^2} &=& \frac{a_1}{\alpha}\rho+\frac{\gamma}{\alpha}a_2\rho^2,\label{F18a}\\
\frac{b(r)}{r^3} &=& \frac{a_3}{\alpha}\rho+\frac{\gamma}{\alpha}a_4\rho^2,\label{F18b}\\
\frac{b^{\prime}(r)}{r^2}-\frac{b(r)}{r^3} &=& \frac{a_5}{\alpha}\rho+\frac{\gamma}{\alpha}a_6\rho^2,\label{F18c}
\end{eqnarray}
where
\begin{eqnarray}
a_1 &=& 1+ \frac{\beta}{2}+\frac{5}{6}\beta\omega_r+\frac{5}{3}\beta\omega_t, \nonumber\\
a_2 &=& \frac{3}{2}-\frac{7}{6}\omega_r^2-\frac{14}{3}\omega_t^2-\frac{14}{3}\omega_r\omega_t-\frac{\omega_r}{3}-\frac{2}{3}\omega_t, \nonumber \\
a_3 &=& -\omega_r-\frac{\beta}{6}\omega_r-\frac{\beta}{2}+\frac{5\beta}{3}\omega_t, \nonumber \\
q_4 &=& -\frac{1}{2} +\frac{5}{6}\omega_r^2-\frac{14}{3}\omega_t^2-\frac{2}{3}\omega_r\omega_t-\frac{\omega_r}{3}+\frac{10}{3}\omega_t,\nonumber\\
a_5 &=& -\omega_t+\frac{2\beta}{3}\omega_t-\frac{\beta}{2}+\frac{5}{6}\omega_r, \nonumber \\
a_6 &=&-\frac{1}{2} +\frac{7}{6}\omega_r^2-\frac{2}{3}\omega_t^2-\frac{8}{3}\omega_r\omega_t+\frac{5\omega_r}{3}+\frac{4}{3}\omega_t.  \label{F20a}
\end{eqnarray}

 Applying the same procedure for $L_m = \rho$ and $L_m = -T$ yields field equations that follow the general forms (ref. \ref{F18a}–\ref{F18c}), though the constants $a_i$ differ significantly. For instance, the assumption $L_m = \rho$ results in
  \begin{eqnarray}
a_1 &=& 1- \frac{\beta}{2}+\frac{\beta}{2}\omega_r+\beta\omega_t, \nonumber\\
a_2 &=& -\frac{1}{2}-\frac{1}{2}\omega_r^2-2\omega_t^2-2\omega_r\omega_t+\omega_r+2\omega_t, \nonumber \\
a_3 &=& -\frac{3}{2}\beta-(\frac{2+\beta}{2})\omega_r+\beta\omega_t, \nonumber \\
a_4 &=& -\frac{5}{2} +\frac{3}{2}\omega_r^2-2\omega_t^2+2\omega_r\omega_t+\omega_r+6\omega_t,\nonumber\\
a_5 &=& -2\omega_t-3\beta+ \beta\omega_r, \nonumber \\
a_6 &=&-5 -\omega_r^2+ 4\omega_t^2+ 6\omega_r +8\omega_t.  \label{F21a}
\end{eqnarray}
This demonstrates that wormhole solutions are sensitive to the choice of the matter Lagrangian density.  A more thorough discussion of the explicit forms of these solutions and their physical properties is necessary. Therefore, expanding the current analysis to include non-linear $f(Q,T)$ models represents a promising avenue for future research and will be addressed in a subsequent publication.

\section{Discussion}\label{sec4}
In this Section, we present a comparative analysis of the results obtained for the different options of $L_m$. It was noted that Eq.( \ref{1cc}) is independent of the selection of $L_m$. However, examining wormhole solutions for various forms of $L_m$ reveals that the choice of $L_m$ explicitly influences the wormhole solutions and their physical properties. The general structure of the field equations is consistent across the three different choices of $L_m$. The distinctions arise from the values of the constants $A_i$ in each case. The similarity in the general structure of the field equations gives rise to a common form for the shape function across the various algorithms used to identify wormhole solutions in this research. For instance, in the case of a linear EoS, the shape function for each of the three different choices of $L_m$ takes the form of a power-law: $b(r)=r^{n_i}$. The distinctions among these cases stem from the dependence of $n_i$ on the parameters $\omega$ and $ \beta$. The relationship between $n_i$ and the parameters $\beta$ and $\omega$ has a direct impact on the range of values for $\beta$ and $\omega$ that allow for the existence of non-exotic wormhole solutions (see Eqs.(\ref{22a} and (\ref{5c})). The findings pertaining to the linear EoS are compiled in Table(\ref{Tab3}). The table illustrates that a particular shape function can result from different combinations of the parameters $\beta$ and $\omega$. It follows that a single shape function may satisfy ECs for one or two selected Lagrangian matter densities, while violating them in other cases. A second point is that a common choice for the parameters $\alpha$, $\beta$, and $\omega$ can lead to different shape functions, and consequently, different outcomes regarding the validity of the ECs.

\begin{table*}[]
\begin{tabular}{|l|l|l|l|l|l|}
\hline
$L_m$  & $b(r)=r^n$ & $\alpha$  & $\beta$ & $\omega$ &  Satisfied ECs   \\ \hline
 $-P$  &  \textcolor{red}{$r^{2/3}$} &-1 & 1 & $\frac{1}{2}$ &  \\ \hline
 $-T$  & \textcolor{red}{$r^{2/3}$} & -1 & $-\frac{2}{5}$ & $-\frac{1}{2}$ &   \\ \hline
$\;\rho$ &\textcolor{red}{$r^{2/3}$} & -2 & 1 & $\frac{1}{2}$ & everyone \\ \hline
 $-P$  &  $r^{-1/4}$ & \textcolor{blue}{-1} & \textcolor{blue}{$\frac{1}{4}$} & \textcolor{blue}{2} & \\ \hline
 $-T$  & $r^{-7/11}$ & \textcolor{blue}{-1} & \textcolor{blue}{$\frac{1}{4}$} & \textcolor{blue}{2} &  \\ \hline
$\;\rho$ & $r^{-1/5}$ & \textcolor{blue}{-1} & \textcolor{blue}{$\frac{1}{4}$} & \textcolor{blue}{2} &  NEC, WEC, SEC\\ \hline
 $-P$  &  $r^{-36/55}$ & \textcolor{green}{-1} & \textcolor{green}{$\frac{1}{5}$} & \textcolor{green}{$\frac{3}{4}$} & \\ \hline
 $-T$  & $r^{-32/17}$ & \textcolor{green}{-1} & \textcolor{green}{$\frac{1}{5}$} & \textcolor{green}{$\frac{3}{4}$} &  everyone \\ \hline
$\;\rho$ & $r^{-12/23}$ & \textcolor{green}{-1} & \textcolor{green}{$\frac{1}{5}$} & \textcolor{green}{$\frac{3}{4}$} & everyone \\ \hline
 \end{tabular}
\caption{The results of ECs for some individual $L_m$ while shape function is identical (red colour) and with the same parameters like $\alpha, \beta,$ and $\omega$ (blue and green).  }\label{Tab3}
\end{table*}

For the variable EoS, the outcomes are consistent with those of the linear EoS. In this instance, when the EoS remains the same across all scenarios, the general structure of the shape functions is alike; however, the influence of parameters such as $B_1, B_2, D_1$ and $D_2$ within the shape functions (refer to the shape functions (\ref{33b}) and (\ref{11c})) directly impacts the relationship between physical properties like $\beta$ and $\omega_\infty$ and the shape function. The results summary for the two selected choices of $\omega(r)$ is presented in Table(\ref{Tab4}). This table illustrates that the range in which ECs are satisfied is influenced by the selection of $L_m$. Table (\ref{Tab4}) shows that the same EoS can result in different shape functions. Conversely, the same shape function may originate from different EoS. It is also apparent that the behavior of the ECs varies in each case. It is evident that the shape function ( \ref{35b}) for $L_m=-T$ resembles ( \ref{16c}) for $L_m=\rho$, although $\beta$ varies in each instance. Therefore, we can deduce that certain identical shape functions may emerge with differing parameter values and also exhibit distinct physical properties.
This result indicates that, depending on the choice of $L_m$ in $f(Q,T)$ gravity, fluids with distinct physical properties can give rise to the same geometry. In other words, a wormhole structure may be formed by different types of fluids. Moreover, the behavior of the ECs varies for each matter Lagrangian density for identical geometry.

\begin{table*}[]
\begin{tabular}{|l|l|l|l|l|l|l|}
\hline
$L_m$  & $b(r)$ & $\alpha$  & $\beta$  & $\rho$ & $\omega(r)$ & Satisfied ECs   \\ \hline
 $-P$  &  \textcolor{red}{$(\frac{2}{r+1})^4$} & -2 & 2  &$\frac{128}{27r^2(r+1)^5}$ & $\frac{41}{4}+\frac{9}{4r}$ & NEC, WEC, SEC \\ \hline
 $-T$  &  \textcolor{red}{$(\frac{2}{r+1})^4$} & -1 & 1  & $\frac{640}{21r^2(r+1)^5}$ & $\frac{3}{4}+\frac{7}{20r}$ &  NEC, WEC, SEC \\  \hline
$\;\rho$ & \textcolor{red}{$(\frac{2}{r+1})^4$} & -2 & 2   & $-\frac{128}{9r^2(r+1)^5}$ &  $-\frac{19}{4}-\frac{3}{4r}$ & \\ \hline
 $-P$  & \textcolor{blue}{$\exp(\frac{1-r^4}{2})$} &-2 & 2 & $\frac{4r}{27}\exp(\frac{1-r^4}{2})$ &  $8+\frac{9}{2r^4}$ &  NEC, WEC, SEC \\ \hline
 $-T$  & \textcolor{blue}{$\exp(\frac{1-r^4}{2})$} & -2 & 2 & $\frac{20r}{21}\exp(\frac{1-r^4}{2})$ & $\frac{2}{5}+\frac{7}{10r^4}$ &  NEC, WEC, SEC  \\ \hline
$\;\rho$ &\textcolor{blue}{$\exp(\frac{1-r^4}{2})$} & -2 & 2 & $-\frac{4r}{9}\exp(\frac{1-r^4}{2})$ & $-4-\frac{3}{2r^4}$ & \\ \hline
 $-P$  & \textcolor{yellow}{$\left(\frac{5}{4r+1}\right)^{1/2}$} & -2 &  $\frac{5}{6}$ & $-\frac{16\sqrt{5}}{11r^2(4r+1)^{3/2}}$ &  $-\frac{11}{4}-\frac{3}{4r}$ &  \\ \hline
 $-T$  & \textcolor{yellow}{$\left(\frac{5}{4r+1}\right)^{1/2}$} & -2 &  $\frac{1}{3}$ &$\frac{5\sqrt{5}}{2r^2(4r+1)^{3/2}}$  & $\frac{13}{5}+\frac{3}{5r}$ & \\ \hline
$\;\rho$ & \textcolor{yellow}{$\left(\frac{5}{4r+1}\right)^{1/2}$} & -1 &  $\frac{1}{4}$ & $\frac{16\sqrt{5}}{5r^2(4r+1)^{3/2}}$ & $\frac{1}{2}+\frac{1}{4r}$ &  everyone \\ \hline
 $-P$  &  $\frac{23^{2/3}}{(28r+5)^{2/3}}$ & \textcolor{green}{-1} & \textcolor{green}{$-\frac{1}{4}$} & $\frac{16*23^{2/3}}{r^2(28r+5)^{2/3}}$ &\textcolor{green}{$\frac{1}{2}+\frac{1}{4r}$} & everyone \\ \hline
 $-T$  & $\frac{343\sqrt{7}}{(4r+3)^{7/2}}$ & \textcolor{green}{-1} &  \textcolor{green}{$-\frac{1}{4}$} & $\frac{16464\sqrt{7}}{5r^2(4r+3)^{9/2}}$ & \textcolor{green}{$\frac{1}{2}+\frac{1}{4r}$} & everyone\\ \hline
$\;\rho$ & $\left(\frac{5}{4r+1}\right)^{1/2}$ & \textcolor{green}{-1} &   \textcolor{green}{$-\frac{1}{4}$} & $\frac{16}{\sqrt{5}r^2(4r+1)^{3/2}}$ &\textcolor{green}{$\frac{1}{2}+\frac{1}{4r}$} & everyone \\ \hline
 \end{tabular}
\caption{The results of the ECs for several choices of $L_m$ are presented for cases in which the shape function is identical (red, blue, yellow), as well as for the case with identical EoS and identical values of $\alpha$ and $\beta$ (green). For the same shape function, different EoS and different energy densities emerge. In contrast, in the green case the identical EoS leads to different shape functions.}\label{Tab4}
\end{table*}

We have analyzed a non-linear form of $f(Q,T)$ gravity that yields second-order field equations. While further rigorous calculation is required to determine the precise analytical wormhole solutions, the structure of the field equations indicates that these solutions are fundamentally dependent on the choice of the matter Lagrangian.

\section{Concluding remarks}\label{sec5}

Wormhole solutions within the framework of $f(Q,T)$ gravity signify a promising field of study that connects theoretical physics with speculative ideas regarding the Universe. Through the investigation of these solutions, scientists seek to enhance our comprehension of gravity, spacetime, and the possibilities of remarkable phenomena that question our traditional perceptions of reality. The exploration of wormholes in this altered context paves the way for new research opportunities, encouraging both theoretical examination and observational investigation into the essence of the cosmos.

The selection of the matter Lagrangian density is a crucial element in modified gravity theories, which greatly influences the resulting field equations and their physical consequences. By thoughtfully choosing or altering the matter Lagrangian, researchers are able to investigate a broad spectrum of gravitational phenomena, assess the feasibility of unconventional solutions, and tackle unresolved issues in cosmology and astrophysics. This adaptability in the formulation of matter interactions not only enriches the theoretical framework but also deepens our comprehension of the fundamental characteristics of gravity and its influence on the Universe's structure. As modified gravity theories progress, the examination of matter Lagrangian will continue to be a central focus, steering the quest for a more profound understanding of gravitational dynamics and the cosmos.

Recently, we have examined non-exotic wormhole solutions within the framework of $f(Q,T)$ gravity for $L_m=-P$ \cite{fq-sara}. The exotic matter problem is resolved in the context of $f(Q,T)$ gravity. In the current paper, we have considered two additional potential forms for the matter Lagrangian, $L_m=\rho$ and $L_m=-T$. In $f(Q,T)$, the gravitational action explicitly contains $T$. Since $T$ is derived from $L_m$, the gravitational field equations are no longer just a function of the metric and curvature/non-metricity, but are explicitly sensitive to the functional form of $L_m$. We demonstrate that the general relation between the radial and lateral NEC in $f(Q,T)$
 gravity and GR, as presented in Eq. (\ref{1cc}), is independent of the chosen matter Lagrangian density. Consequently, solutions that simultaneously violate the radial and lateral NEC in the context of GR may satisfy these conditions within $f(Q,T)$ gravity.

 It was shown that the general form of the field equations is identical for different matter Lagrangian densities. However, the relations between the different $L_m$ choices and their corresponding physical properties arise through the connection mediated by the parameters $\alpha$ and $\beta$ in $A_i$. The linear EoS leads to a power-law shape function, a particularly well-known form in wormhole physics. When the EoS is generalized to an asymptotically linear form, the range of admissible solutions becomes significantly wider. The resulting wormhole configurations depend primarily on the parameters $\alpha$ and $\beta$, which govern the dynamics of the model. Meanwhile, other parameters such as$n_i$, $\omega$, $D$, $n$, $B_i$ and $D_i$, appearing in the shape functions, determine the geometric attributes of the spacetime. The coefficients $A_i$ act as mediators linking the key coupling parameters to the geometric characteristics for each choice of matter Lagrangian density. This analysis reveals a variety of physical behaviors among the solutions, which represents the core outcome of the present investigation.

This work demonstrates the accessibility of non-exotic, asymptotically flat wormhole solutions within the context of three common matter Lagrangian densities ($L_m$) found in the literature. We have established that the ECs are not universally satisfied for a given shape function across all $L_m$  choices; indeed, a shape function may admit the ECs for one specific $L_m$ while violating them for others. Our results further reveal a duality: the same shape function can be supported by disparate fluid configurations, or conversely, identical fluids can manifest in different geometries. Such results underscore the pivotal role of the matter Lagrangian density in modified gravity theories, exemplified by $f(Q,T)$ gravity. In modified gravity theories like $f(Q,T)$, the choice of $L_m$ is no longer a purely mathematical starting point. It acts as a physical parameter that dictates how matter "back-reacts" on the geometry. This forces physicists to adopt more rigorous definitions for $L_m$ to ensure the resulting theory is physically viable and avoids instabilities. Understanding the role of the matter Lagrangian density in modified theories is crucial for exploring their viability, studying cosmological solutions, and comparing predictions with observations. This understanding also sheds light on how gravitational modifications could mimic or explain dark energy phenomena without invoking unknown exotic components.


\begin{thebibliography}{99}






\bibitem{Visser} M. Visser,\textit{ Lorentzian wormholes: From Einstein to Hawking}, (AIP Press, New York, 1995).
 \bibitem{cut} M. Visser, S. Kar and N. Dadhich, Phys. Rev. Lett.  {\bf 90}, 201102 (2003).

 \bibitem{cut2} N. M. Garcia, F. S. N. Lobo, and M. Visser, Phys. Rev. D {\bf 86}, 044026 (2012).

\bibitem{cut3}S. D. Forghani, S. H. Mazharimousavi, and M. Halilsoy, Phys. Lett. B {\bf804},  135374 (2020).
\bibitem{variable} F. Parsaei and S. Rastgoo, Phys. Rev. D {\bf 99}, 104037 (2019).

 \bibitem{foad} F. Parsaei and S. Rastgoo,  Eur. Phys. J. C  {\bf 80}, 366 (2020).

\bibitem{phantom} F. S. N. Lobo, Phys. Rev. D {\bf 71}, 084011 (2005).
\bibitem{phantom2}J.A. Gonzalez, F. S. Guzman, N. Montelongo-Garcia, and T. Zannias, Phys. Rev. D {\bf 79}, 064027 (2009).
\bibitem{phantom1}F. S. N. Lobo, F. Parsaei, and N. Riazi, Phys. Rev. D {\bf 87}, 084030 (2013).





\bibitem{casm}M. R. Mehdizadeh, and A. H. Ziaie, Eur. Phys. J. Plus. {\bf 139}, 1001 (2024).


  \bibitem{b1} K. A. Bronnikov and Sung-Won Kim, Phys. Rev. D {\bf67}, 064027 (2003).
 \bibitem{b2}F. Parsaei, N. Riazi,  Phys.\ Rev.\ D {\bf 91}, 024015 (2015).
 \bibitem{b3} F. Parsaei, N. Riazi,  Phys.\ Rev.\ D {\bf 102}, 044003 (2020).

\bibitem{Bo}  M. G. Richarte, C. Simeone, Phys. Rev. D {\bf 80}, 104033 (2009).
 \bibitem{Bo1} R. Shaikh, Phys. Rev. D { \bf 98} , 064033 (2018).

\bibitem{quad}F. Duplessis, and D. A. Easson, Phys. Rev. D {\bf 92}, 043516 (2015).
 \bibitem{quad1} H. K. Nguyen , and M. Azreg-Aïnou, Eur. Phys. J. C {\bf 83}, 626 (2023).

\bibitem{Cartan} K. A. Bronnikov and A. M. Galiakhmetov, Grav. Cosmol {\bf21}, 283 (2015).
 \bibitem{Cartan1} M.R. Mehdizadeh and A.H. Ziaie, Phys. Rev. D {\bf 95}, 064049 (2017).
 \bibitem{Cartan2} E. D. Grezia, E. Battista, M. Manfredonia, and G. Miele Eur. Phys. J. Plus. {\bf 132}, 537 (2017).
 \bibitem{Cartan3} S. V. Soni, A. C. Khunt, and A. H. Hasmani, Int. J. Geom. Methods. Mod. Phys {\bf06}, 2450115 (2024).
 \bibitem{Rast} H. Moradpour, N. Sadeghnezhad and H. Hendi, Can. J. Phys. { \bf 95}, 1257 (2017).
 \bibitem{Rast1} F. Parsaei, S. Rastgoo,  Phys. Lett. B {\bf874}, 140276 (2026).
 \bibitem{Sadegh} N. Sadeghnezhad, Int. J. Geom. Methods. Mod. Phys {\bf02}, 2550153 (2025).
 \bibitem{fq} F. Parsaei, S. Rastgoo, and P. K. Sahoo Eur. Phys. J. Plus {\bf 137}, 1083 (2022).
\bibitem{fq1} M. Calzá, and L. Sebastiani, Eur. Phys. J. C {\bf 83}, 247 (2023).
\bibitem{fq2} S. Rastgoo, and F. Parsaei,  Eur. Phys. J. C  {\bf 84}, 563 (2024).
\bibitem{fq4} A. Banerjee, A. Pradhan, T. Tangphati, and F. Rahaman, Eur. Phys. J. C {\bf 81}, 1031 (2021).
\bibitem{fq44} Z. Hassan, S. Mandal, and P.K. Sahoo, Fortschr. Phys. {\bf69}, 2100023 (2021).
\bibitem{fq5}S. Kiroriwal, J. Kumar, S.K. Maurya, and S. Ray, Phys. Dark Universe {\bf46}, 101559 (2024).
\bibitem{Nojiri}S.~Nojiri, S.~D.~Odintsov, and V.~K.~Oikonomou, Phys. Rept. {\bf692}, 1 (2017).
\bibitem{fR11} E. F. Eiroa, and G. F. Aguirre,  Eur. Phys. J. C  {\bf 6}, 132 (2016).
\bibitem{fR22} T. Multam¨aki, I. Vilja, Phys. Rev. D 74, 064022 (2006).
\bibitem{fR55}A. S. Agrawal, B. Mishra, F. Tello-Ortiz, and A. Alvarez, Fortschr. Phys. {\bf70}, 2100177 (2022).

\bibitem{Must1} G. Mustafa, M. Ahmad, A. Övgün, M. F. Shamir, and I. Hussain, Fortschr. Phys. {\bf61} ,2100048 (2021).
\bibitem{Err}	A. Errehymy , S. Hansraj, S.K. Maurya, C. Hansraj a, M. Daoud,  Phys. Dark Universe {\bf41}, 101258 (2023).
\bibitem{Riz} M. M. Rizwan, Z. Hassan,P. K. Sahoo, and A. Övgün, Eur. Phys. J. C {\bf 84}, 1132 (2024).
\bibitem{Must2} G. Mustafa, A. Errehymy, F. Javed, S.K. Maurya, S. Hansraj, S. Sadiq, J. High Energy Astrophys. {\bf42}, 1 (2024).
\bibitem{foad3}   F. Parsaei, and S. Rastgoo, Ann. Phys. {\bf482}, 170205 (2025).



\bibitem{L1}A. Errehymy , S.K. Maurya, S. Hansraj, M. Mahmoud, K. S. Nisar, and A. Abdel-Aty,  Chin. J. Phys.  {\bf89}, 56 (2024).
\bibitem{L2} T. Naseer, M. Sharif, M. Faiza,  Chin. J. Phys.  {\bf94}, 204 (2025).

\bibitem{L7}M. Khatri, Z. Chhakchhuak, and A. Lalchhuangliana, Ann. Phys. {\bf470},  169788 (2024).
\bibitem{L10}G. Mustafa, F. Javed, S. K . Maurya, M. Govender, and A. Saleem, Phys. Dark Universe {\bf45},   101508 (2024).
\bibitem{L12} M. Z. Rizwan, Z. Hassan, and P.K. Sahoo Phys. Lett. B {\bf860}, 139152 (2025).
\bibitem{SR2} S. Rastgoo, and F. Parsaei, Ann. Phys. {\bf488}, 170394 (2026).

\bibitem{Azizi} T. Azizi, Int. J. Theor. Phys, {\bf 52} 3486 (2013).
\bibitem{Moa} P. H. R. S. Moraes, and P. K. Sahoo, Phys. Rev. D {\bf 96}, 044038 (2017).
\bibitem{Zub} M. Zubair, G. Mustafa, S. Waheed, and G. Abbas,  Eur. Phys. J. C {\bf 77}, 680 (2017).
\bibitem{Zub2}M. Zubair, R. Saleem, Y. Ahmad, and G. Abbas,  Int. J. Geom. Methods. Mod. Phys {\bf16}, 1950046 (2019).





\bibitem{fr2} E. Elizalde and M. Khurshudyan, Phys. Rev. D {\bf 98}, 123525 (2018).
\bibitem{Sha} U. k. Sharma, abd A. K. Mishra,  Found. Phys {\bf51}, 50 (2021).


\bibitem{Ban}A. Banerjee, M.K. Jasim, and S. G. Ghosh, Ann. Phys, {\bf433}, 16875 (2021).
\bibitem{Sarkar} N. Sarkar, S. Sarkar, A. Bouzenada, A. Dutta, M. Sarkar, and F. Rahaman, Phys. Dark Universe {\bf44}, 101439 (2024).
\bibitem{SR1} S. Rastgoo, and F. Parsaei, Nucl. Phys. B {\bf1011}, 116797 (2025).
\bibitem{foad4}  F Parsaei and S Rastgoo,  Commun. Theor. Phys. {\bf78}, 025403 (2026).











\bibitem{1} N. S. Kavya, G. Mustafa, and V. Venkatesha, Ann. Phys, {\bf468},  169723  (2024).
\bibitem{11}C. C. Chalavadi., N.S. Kavya, V. Venkatesha,  Eur. Phys. J. Plus {\bf 138}, 885 (2023).
\bibitem{12a}C. C. Chalavadi, V. Venkatesha, N. S. Kavya, and S. V. D. Rashmi, Commun. Theor. Phys. {\bf76}, 025403 (2024).
\bibitem{12} M. Tayde, Z. Hassan, P.K. Sahoo, Chin. J. Phys. {\bf89},  195 (2024).
\bibitem{2} G. G. L. Nashed, W. E. Hanafy, JCAP {\bf09}, 040 (2025).
\bibitem{3} M. Tayde, Z. Hassan, P.K. Sahoo, Nucl. Phys. B {\bf1000}, 116478 (2024).
\bibitem{4} S. D. Sadatian, and S. M. R. Hosseini,  Adv. High Energy Phys. {\bf2024},  3717418 (2024).
\bibitem{5} M. Tayde, Z. Hassan, P.K. Sahoo, and S. Gutt, Chin. Phys. C {\bf46}, 115101 (2022).
\bibitem{6} M. Tayde, J. R. L. Santos, J. N. Araujo, and P. K. Sahoo, Eur. Phys. J. Plus {\bf 138}, 539 (2023).
\bibitem{7} M. Tayde, S. Ghosh, and P.K. Sahoo, Chin. Phys .C  {\bf47}, 075102 (2023).
\bibitem{8} V. Venkatesha, C. C. Chalavadi, N.S. Kavya, and P.K. Sahoo,  New Astron. {\bf105},  102090 (2024).
\bibitem{9} M. Tayde, Z. Hassan, P.K. Sahoo, Phys .Dark Univ. {\bf42},  101288 (2023).
\bibitem{10} A. Sahoo, S. K. Tripathy, B. Mishra, and S. Ray, Eur. Phys. J. C {\bf84}, 325 (2024).


\bibitem{13} M. Zeeshan Gul, M Sharif, S. Shahid, and F. Javed,  Phys. Scr. {\bf99}, 125004 (2024).
\bibitem{14} A. Pradhan, M. Zeyauddin, A. Dixit, and K. Ghaderi, Universe {\bf11}, 279 (2025).

\bibitem{fq-sara} S. Rastgoo, F. Parsei, and S. Nasirimoghadam, arXiv:2602.00527v1.

\bibitem{fqt}Y. Xu et al., Eur. Phys. J. C, {\bf79}, 708 (2019).
\bibitem{20} P.H.R.S. Moraes, Eur. Phys. J. C, {\bf79}, 674 (2019).
\bibitem{Haghani} Z. Haghani, T. Harko, and S. Shahidi, Phys .Dark Univ. {\bf44},  101448 (2024).
\bibitem{21} T. P. Sotiriou and V. Faraoni, Class. Quant. Grav. {\bf25}, 205002 (2008).
\bibitem{22} O. Bertolami, F. S. N. Lobo and J. Paramos, Phys. Rev. D {\bf78}, 064036 (2008)
\bibitem{23} T. Harko, F. S. N. Lobo, S. Nojiri, and S. D. Odintsov, Phys. Rev. D {\bf84}, 024020 (2011).
\bibitem{24} S. Mendoza, and S. Silva, Int. J. Geom. Meth. Mod. Phys. {\bf23}, 2550211 (2026).
\bibitem{25} D. Bhattacharjee, and P. K. Chattopadhyay,  Phys. Lett. B {\bf873}, 140138 (2026).
\bibitem{26} R. Bhagat, I. V. Fomin, and B. Mishra, arXiv:2509.13382v1.
\bibitem{26b} P.P. Avelino and R.P.L. Azevedo, Phys. Rev. D {\bf97}  064018 (2018).
\end{thebibliography}
\end{document}